\documentclass[preprint, showpacs, aps, pra]{revtex4-1}	
\usepackage{latexsym,amssymb,makeidx}

\usepackage{epsfig, color, ulem}
\usepackage{graphicx}
\usepackage{dcolumn}
\usepackage{bm}
\usepackage{amsmath}
\usepackage{setspace}

\begin{document}
\bibliographystyle{jasanum}

\title{On the relation between time-reversed acoustics and Green's function retrieval in space-variant and in time-variant materials}
\author{Kees Wapenaar$^a$, Johannes Aichele$^b$, Dirk-Jan van Manen$^b$}
\affiliation{$^a$Delft University of Technology, Department of Geoscience and Engineering, Stevinweg 1, 2628 CN Delft, The Netherlands \\
$^b$ETH Z\"urich, Institute of Geophysics, Sonneggstrasse 5, 8092 Z\"urich, Switzerland}
\date{\today}

\maketitle

\begin{spacing}{1.}

\centerline{\Large Abstract}

\noindent 

The methods of time-reversed acoustics and Green's function retrieval  are traditionally deployed for classical inhomogeneous, time-invariant materials. 
The mutual relation between these methods is well-established. Recently, similar methods have been proposed for homogeneous, time-variant materials. 
Here we investigate their mutual relation and their relation with the corresponding methods in classical materials. 
For this analysis we make use of the fact that the wave equations for both classes of material are similar, 
with the roles of time and space interchanged. However, the principle of causality holds for both classes of material, 
hence, here the roles of time and space are not interchanged. We find that
\begin{itemize}
\item whereas classical time-reversed acoustics involves emission of a time-reversed single-component wave field from a (ideally closed) 
boundary into the inhomogeneous material, its idealized counterpart involves emission of a sign-reversed two-component wave field, 
recorded in a time-reversed material, from a single time instant into the actual time-variant material;
\item whereas classical Green's function retrieval involves temporal crosscorrelation of wave fields at two space locations 
in response to single-component sources on a (ideally closed) boundary, its counterpart involves spatial crosscorrelation of wave fields 
at two time instants in response to two-component sources at a single time instant.
\end{itemize}


\section{Introduction}

Waves propagating through time-variant materials are scattered by time boundaries, 
just as waves propagating through space-variant materials are scattered by spatial boundaries.
For electromagnetic waves, scattering by time boundaries has been studied for decades \cite{Morgenthaler58IRE, Jiang75IEEE, Mendonca2002PS}.
The research in this area expanded significantly since the introduction of dynamic metamaterials
\cite{Koutserimpas2018IEEE, Caloz2020IEEE1, Ramaccia2020OL,  Apffel2022PRL, Moussa2023NP, Pacheco2025PRB, Stefanini2025PRL}.
For acoustic waves, the phenomenon of time scattering has been introduced by Fink and coworkers 
\cite{Bacot2016NP, Fink2017EPJ,  Bal2019SIAM, Hidalgo2023PRL, Delory2025PRL}.
In particular, they demonstrated theoretically and experimentally that a time boundary sends waves back in space to focus at their origin, 
similar to time-reversed waves that are sent back from a spatial boundary \cite{Fink92IEEE}.

Whether we consider acoustic or electromagnetic waves, there exists much analogy between wave phenomena in homogeneous, time-variant materials,
and those in ``classical'' inhomogeneous, time-invariant materials 
\cite{Mendonca2002PS, Xiao2014OL, Hoop2014WM, Salem2015arXiv, Torrent2018PRB, Caloz2020IEEE2, Manen2024arXiv, Wapenaar2024WM}. 
In particular, the roles of time and space are interchanged between the wave equations for both classes of material.
However, the analogy also has its limitations. 
For example, waves reflected by a time boundary  propagate forward in time and backward in space, similar as waves reflected by a spatial boundary
 \cite{Xiao2014OL, Salem2015arXiv, Caloz2020IEEE2}. Hence, here the roles of  time and space are not interchanged.
More generally, the principle of causality is a time-related property, whether we consider time-variant or space-variant materials \cite{Wapenaar2024WM}.
In other words, whereas the roles of time and space are interchanged between the wave equations for both classes of material, 
they are in general not interchanged in the initial and boundary conditions, 
and hence the relation between the solutions of the wave equations (i.e., the wave fields in both classes of material)  is in general not straightforward.
Another aspect that complicates the analogy is the fact that the number of time dimensions (one) is different from the number of space dimensions (two or three), 
except of course for 1D situations. 

Despite the aforementioned limitations, we investigate the relation between time-reversed acoustics and Green's function retrieval in space-variant and in time-variant materials.
For classical inhomogeneous, time-invariant materials, the research fields of time-reversed acoustics \cite{Fink92IEEE, Draeger99JASA, DeRosny2002PRL}
and Green's function retrieval by temporal crosscorrelation 
\cite{Weaver2001PRL, Campillo2003Science, Wapenaar2003GEO, Snieder2004PRE, Schuster2004GJI, VanManen2005PRL} 
originated independently. However, it was soon
discovered that these fields are closely related \cite{Derode2003JASA, Wapenaar2005JASA}. It is reasonable to assume that a similar close relationship exists 
between time-reversed acoustics \cite{Bacot2016NP, Fink2017EPJ}
and Green's function retrieval by spatial crosscorrelation \cite{Wapenaar2024WM, Wapenaar2025RS} in homogeneous, time-variant materials.
In a companion paper \cite{Aichele2025arXiv} we investigate the relations between temporal and spatial crosscorrelation, 
underlying time-reversed acoustics and Green's function retrieval. 
In that analysis we keep the material space- and time-invariant, except for an impulsive temporal disruption, representing a time boundary. 
In this paper we extend the analysis for materials that are either arbitrarily space-variant or arbitrarily time-variant.

After a brief discussion of the general equations for waves in space- and time-variant materials (section \ref{sec2}), in section \ref{sec3}
we review classical time-reversed acoustics, Green's function retrieval and their mutual relation for arbitrarily inhomogeneous, time-invariant materials.
This serves as an introduction to section \ref{sec4}, where
we discuss time-reversed acoustics, Green's function retrieval and their mutual relation for homogeneous, arbitrarily time-variant materials.
In that section we also discuss the relation with the classical methods reviewed in section \ref{sec3}. We present conclusions in section \ref{sec5}.

\section{General equations for waves in space- and time-variant materials}\label{sec2}

Although in this paper we deal primarily with 2D and 3D acoustic waves, we use a unified notation that also covers 2D horizontally-polarized shear (SH) waves,
2D transverse electric (TE) and 2D transverse magnetic (TM) waves (the latter two are relevant because much of the theory for waves in time-variant materials
was initially developed for electromagnetic waves).
We denote position in space by the Cartesian coordinate vector ${\bf x}=(x,y,z)$ (for the 3D situation) or ${\bf x}=(x,z)$ (for the 2D situation), and time by $t$.
With reference to Table \ref{table1}, we consider space- and time-variant wave field quantities $U({\bf x},t)$, ${\bf V}({\bf x},t)$, $P({\bf x},t)$ and ${\bf Q}({\bf x},t)$,
space- and time-variant material parameters $\alpha({\bf x},t)$ and $\beta({\bf x},t)$, and space- and time-variant source quantities $a({\bf x},t)$, ${\bf b}({\bf x},t)$.
The boldface quantities ${\bf V}$, ${\bf Q}$ and ${\bf b}$ denote vectors with three components, i.e. ${\bf V}=(V_x,V_y,V_z)$ etc. (for the 3D situation),
or with two components, i.e. ${\bf V}=(V_x,V_z)$ etc. (for the 2D situation).
The wave field quantities obey the following unified equations
\begin{eqnarray}
\partial_t U + {\bf \nabla}\cdot {\bf Q} &=& a,\label{eq1}\\
\partial_t {\bf V} + {\bf \nabla} P &=& {\bf b},\label{eq2}
\end{eqnarray}
where $\partial_t$ stands for the partial differential operator $\frac{\partial}{\partial t}$, and ${\bf \nabla}$ is the nabla operator,
defined as ${\bf \nabla}=(\partial_x,\partial_y,\partial_z)$ (for the 3D situation) or ${\bf \nabla}=(\partial_x,\partial_z)$ (for the 2D situation).
For acoustic waves, using the quantities defined in row 1 of Table \ref{table1}, 
equation (\ref{eq1}) is the acoustic deformation equation,  $-\partial_t \Theta + {\bf \nabla}\cdot {\bf v} = q$, and 
equation (\ref{eq2})  stands for equilibrium of mechanical momentum, $\partial_t {\bf m} + {\bf \nabla}p = {\bf f}$ \cite{Hoop95Book, Torrent2018PRB}.

Assuming the material is isotropic and instantaneously and locally reacting, 
the unified wave field quantities in equations (\ref{eq1}) and (\ref{eq2}) are related via the  constitutive relations
\begin{eqnarray}
U &=& \alpha P,\label{eq3}\\
{\bf V} &=& \beta {\bf Q}.\label{eq4}
\end{eqnarray}
For acoustic waves, these relations are $-\Theta=\kappa p$ and ${\bf m}=\rho{\bf v}$, respectively.

Equations (\ref{eq1}) -- (\ref{eq4}) hold in materials of which the parameters $\alpha({\bf x},t)$ and $\beta({\bf x},t)$ vary continuously with space and time.
They need to be supplemented with boundary conditions at those positions and times where these parameters are discontinuous.
At a spatial boundary with normal ${\bf n}({\bf x})$, where the parameters are discontinuous in space but continuous in time, the boundary conditions
demand that $P$ and ${\bf Q}\cdot{\bf n}$ are continuous (for acoustic waves, these are the well-known boundary conditions for the acoustic pressure $p$
and the normal component of the particle velocity ${\bf v}\cdot{\bf n}$). 
At a time boundary, where the parameters are discontinuous in time but continuous in space, the boundary conditions
demand that $U$ and ${\bf V}$ are continuous \cite{Xiao2014OL, Morgenthaler58IRE, Mendonca2002PS, Caloz2020IEEE2, Hoop2014WM}.
Several authors discuss customized approaches for materials with specific combinations of spatial and time boundaries 
\cite{Apffel2022PRL, Caloz2020IEEE2, Manen2024arXiv}.
Note that at the intersection of a spatial boundary with a time boundary, the boundary conditions for the spatial boundary are inconsistent with those for the time boundary. 
This can be solved by the introduction of a field-dependent point-source at the intersection \cite{Stefanini2025PRL}.
Other authors discuss ``travelling-wave modulated'' materials, which exhibit non-reciprocal wave behaviour
\cite{Lurie2007Book,  Trainiti2016NJP, Nassar2017JMPS, Nassar2017RS, Goldsberry2019JASA, Sotoodehfar2023OE}.

A further discussion of waves in materials in which the parameters vary with space and time is beyond the scope of this paper.
In the following we consider either  arbitrarily inhomogeneous, time-invariant materials, with  parameters $\alpha({\bf x})$ and $\beta({\bf x})$ (section \ref{sec3}),
or homogeneous, arbitrarily time-variant materials, with parameters $\alpha(t)$ and $\beta(t)$ (section \ref{sec4}).

\begin{table}[h]
\caption{
Specification of the quantities in equations (\ref{eq1}) -- (\ref{eq4}). 
For 2D and 3D acoustic waves, the quantities are cubic dilatation $\Theta$, mechanical momentum density ${m}$, acoustic pressure $p$, particle velocity ${v}$,  
compressibility $\kappa$, mass density $\rho$,  volume-injection rate density $q$ and external force density ${f}$.
For 2D SH (horizontally polarized shear) waves in the $(x,z)$-plane, $m$, $v$, $\rho$ and $f$ are defined the same as for acoustic waves, and 
the additional quantities are strain $e$,  stress $\tau$, shear modulus $\mu$ and external  deformation rate density $h$.
For 2D TE (transverse electric) and 2D TM (transverse magnetic) waves in the $(x,z)$-plane, the
quantities are electric and magnetic flux densities ${D}$ and ${B}$,
electric and magnetic field strengths ${E}$ and ${H}$, permittivity $\varepsilon$, permeability $\mu$,
and external electric and magnetic current densities ${J}^{\rm e}$ and ${J}^{\rm m}$. }\label{table1}
\begin{tabular}
{lcccccccccccccc}
\hline
& $U$ & $V_x$ &$V_y$ &$V_z$ &$P$ & $Q_x$ & $Q_y$ &$Q_z$ &$\alpha$ &$\beta$  & $a$ & $b_x$ & $b_y$ &$b_z$  \\
\hline
1. 2D/3D Acoustic  & $-\Theta$ & $m_x$&$m_y$ &$m_z$& $p$ & $v_x$ & $v_y$ & $v_z$ &$\kappa$ &$\rho$  & $q$ & $f_x$ & $f_y$ &$f_z$ \\
2. 2D SH  & $m_y$ & $-2e_{yx}$ &-& $-2e_{yz}$& $v_y$ & $-\tau_{yx}$ &-& $-\tau_{yz}$ &$\rho$ &$\frac{1}{\mu}$ & $f_y$ & $2h_{yx}$&- &$2h_{yz}$  \\
3. 2D TE   & $D_y$ & $B_z$&-&$-B_x$& $E_y$ & $H_z$ &- & $-H_x$ &$\varepsilon$ &$\mu$  &  $-J_y^{\rm e}$ & $-J_z^{\rm m}$ &- &$J_x^{\rm m}$  \\
4. 2D TM   & $B_y$ & $-D_z$ &- &$D_x$& $H_y$ & $-E_z$ &-&$E_x$ &$\mu$ &$\varepsilon$  & $-J_y^{\rm m}$ & $J_z^{\rm e}$&-&$-J_x^{\rm e}$  \\
\hline
\end{tabular}
\end{table}

\section{Time-reversed acoustics and Green's function retrieval in inhomogeneous, time-invariant materials}\label{sec3} 

In this section we review classical time-reversed acoustics and Green's function retrieval in inhomogeneous, time-invariant materials, and their mutual relation.
In section \ref{sec3A} we discuss the wave equation for an inhomogeneous, time-invariant material, as a special case of the equations in section \ref{sec2}.
In section \ref{sec3B} we discuss the Green's function of an inhomogeneous, time-invariant material. 
In section \ref{sec3H}  we review a classical representation of the homogeneous Green's function of an inhomogeneous, time-invariant material.
In sections \ref{sec3C} and \ref{sec3D} we show how this homogeneous Green's function representation leads to 
time-reversed acoustics and Green's function retrieval, respectively, in inhomogeneous, time-invariant materials.
The review in this section serves as an introduction to section \ref{sec4}, where, in an analogous way, 
we discuss time-reversed acoustics and Green's function retrieval in homogeneous, time-variant materials.

\subsection{Wave equation for an inhomogeneous, time-invariant material}\label{sec3A}

We consider an arbitrarily inhomogeneous, time-invariant material, with  parameters $\alpha({\bf x})$ and $\beta({\bf x})$.
Substituting constitutive relations (\ref{eq3}) and (\ref{eq4}) into equations (\ref{eq1}) and (\ref{eq2}), using the fact that $\alpha({\bf x})$ and $\beta({\bf x})$ are time-invariant,
we obtain
\begin{eqnarray}
\alpha\partial_t P +{\bf \nabla}\cdot{\bf Q} &=& a,\label{eq6}\\
\beta\partial_t {\bf Q} + {\bf \nabla} P &=& {\bf b}.\label{eq7}
\end{eqnarray}
For acoustic waves these expressions become $\kappa\partial_tp + {\bf \nabla}\cdot {\bf v}=q$ and $\rho\partial_t{\bf v}+{\bf \nabla}p={\bf f}$. 
We continue with the general notation of equations (\ref{eq6}) and (\ref{eq7}), so that all that follows also holds for 2D SH, TE and TM waves.

The parameters $\alpha({\bf x})$ and $\beta({\bf x})$ in equations (\ref{eq6}) and (\ref{eq7}) may vary continuously with position.
For a material with piecewise continuous parameters, these equations are supplemented with boundary conditions.
At a spatial boundary with normal ${\bf n}({\bf x})$, where the parameters undergo a finite jump, the boundary conditions
state that $P$ and ${\bf Q}\cdot{\bf n}$ are continuous over that boundary.

By eliminating ${\bf Q}$ from equations (\ref{eq6}) and (\ref{eq7}) we obtain the well-known second order wave equation
\begin{eqnarray}
\biggl(\frac{1}{\beta c^2}\partial_t^2-{\bf \nabla}\cdot\frac{1}{\beta}{\bf \nabla}\biggr)P({\bf x},t)=s({\bf x},t),\label{eq31x}
\end{eqnarray}
with $c({\bf x})$ being the space-variant propagation velocity,  defined as
\begin{eqnarray}
c=\frac{1}{\sqrt{\alpha\beta}}\label{eq8ah}
\end{eqnarray}
and  $s({\bf x},t)$ the source distribution, defined as
\begin{eqnarray}
s=\partial_t a - {\bf \nabla}\cdot \biggl(\frac{1}{\beta}{\bf b}\biggr).\label{eqsourcex}
\end{eqnarray}

\subsection{Green's function of an inhomogeneous, time-invariant material}\label{sec3B}

We introduce the Green's function ${\cal G}_x({\bf x},{\bf x}_A,t,t_A)$ of an inhomogeneous, time-invariant 
material as the response to an impulsive point source at ${\bf x}={\bf x}_A$ at time instant $t=t_A$,
observed at position ${\bf x}$ and time $t$. Hence, it obeys wave
equation (\ref{eq31x}), with the source term $s({\bf x},t)$ on the right-hand side replaced by an impulsive point source, according to
\begin{eqnarray}
\biggl(\frac{1}{\beta c^2}\partial_t^2-{\bf \nabla}\cdot\frac{1}{\beta}{\bf \nabla}\biggr){\cal G}_x({\bf x},{\bf x}_A,t,t_A)=\delta({\bf x}-{\bf x}_A)\delta(t-t_A),\label{eq31xg}
\end{eqnarray}
where $\delta(\cdot)$ is the Dirac delta function. The subscript $x$ in ${\cal G}_x$ denotes that this is the Green's function of a space-variant material.
The Green's function obeys the causality condition
\begin{eqnarray}
{\cal G}_x({\bf x},{\bf x}_A,t,t_A)=0 \quad \mbox{for}\quad t<t_A.\label{eq31cond}
\end{eqnarray}
Assuming that the material is homogeneous outside a sphere with arbitrarily large but finite radius, this causality condition implies that ${\cal G}_x({\bf x},{\bf x}_A,t,t_A)$
is outward propagating for $|{\bf x}-{\bf x}_A|\to\infty$. 

Since the material is time-invariant, the Green's function is time-shift invariant, hence
\begin{eqnarray}
{\cal G}_x({\bf x},{\bf x}_A,t,t_A)={\cal G}_x({\bf x},{\bf x}_A,t-t_A,0).\label{eq31shfd}
\end{eqnarray}

The Green's function obeys the following reciprocity relation \cite{Morse53Book, Hoop95Book}
\begin{eqnarray}
{\cal G}_x({\bf x}_A,{\bf x}_B,t,0)={\cal G}_x({\bf x}_B,{\bf x}_A,t,0).\label{eq31shfrec}
\end{eqnarray}
Hence, in an inhomogeneous, time-invariant material,
the response to a source at ${\bf x}_B$, observed at ${\bf x}_A$, is equal to the response to a source at ${\bf x}_A$, observed at ${\bf x}_B$.

Since wave equation (\ref{eq31xg}) contains only even-order time derivatives, the time-reversed Green's function ${\cal G}_x({\bf x},{\bf x}_A,-t,0)$ is also a solution
of this equation (with $t_A=0$). This acausal Green's function is inward propagating for $|{\bf x}-{\bf x}_A|\to\infty$. It propagates 
through the inhomogeneous material, converges to a sink at ${\bf x}={\bf x}_A$ and $t=0$, after which it disappears.

We define the homogeneous Green's function of an inhomogeneous, time-invariant material as the difference of the Green's function and its time-reversed version, according to
\begin{eqnarray}
{\cal G}_x^h({\bf x},{\bf x}_A,t,0)={\cal G}_x({\bf x},{\bf x}_A,t,0)-{\cal G}_x({\bf x},{\bf x}_A,-t,0).\label{eqghom}
\end{eqnarray}
Both terms on the right-hand side obey the same wave equation, with the same singularity $\delta({\bf x}-{\bf x}_A)\delta(t)$.
The singularities cancel when we subtract the equations for the Green's function and its time-reversed version. Hence, the homogeneous Green's function
obeys the following wave equation
\begin{eqnarray}
\biggl(\frac{1}{\beta c^2}\partial_t^2-{\bf \nabla}\cdot\frac{1}{\beta}{\bf \nabla}\biggr){\cal G}_x^h({\bf x},{\bf x}_A,t,0)=0.\label{eq31xghom}
\end{eqnarray}
Note that ${\cal G}_x^h({\bf x},{\bf x}_A,t,0)$  is called the homogeneous Green's function because it is a solution of this homogeneous differential equation
(hence, here ``homogeneous'' does not refer to the material parameters, which may still be  inhomogeneous, but to the absence of the singularity in the wave equation).
Unlike the Green's function ${\cal G}_x({\bf x},{\bf x}_A,t,0)$ and its time-reversed version ${\cal G}_x({\bf x},{\bf x}_A,-t,0)$, 
the homogeneous Green's function ${\cal G}_x^h({\bf x},{\bf x}_A,t,0)$ is not singular for ${\bf x}\to{\bf x}_A$ and $t\to 0$.

From equations (\ref{eq31shfrec}) and (\ref{eqghom}) it follows that the homogeneous Green's function obeys the following reciprocity relation
\begin{eqnarray}
{\cal G}_x^h({\bf x}_A,{\bf x}_B,t,0)={\cal G}_x^h({\bf x}_B,{\bf x}_A,t,0).\label{eq31shfreckkk}
\end{eqnarray}

\subsection{Homogeneous Green's function representation for an inhomogeneous, time-invariant material}\label{sec3H}

We derive a boundary integral representation for the homogeneous Green's function ${\cal G}_x^h({\bf x},{\bf x}_A,t,0)$
of an inhomogeneous, time-invariant material. We start by defining a vector function ${\bf a}({\bf x}',t)$ as
\begin{eqnarray}
{\bf a}({\bf x}',t)&=&\frac{1}{\beta({\bf x}')}\Bigl[\{{\bf \nabla}'{\cal G}_x({\bf x}',{\bf x}_B,t,0)\}*_t{\cal G}_x({\bf x}',{\bf x}_A,-t,0)
\nonumber\\&&
-{\cal G}_x({\bf x}',{\bf x}_B,t,0)*_t{\bf \nabla}'{\cal G}_x({\bf x}',{\bf x}_A,-t,0)\Bigr],
\end{eqnarray}
where ${\cal G}_x({\bf x}',{\bf x}_A,t,0)$ and ${\cal G}_x({\bf x}',{\bf x}_B,t,0)$ are Green's functions with their sources at ${\bf x}_A$ and ${\bf x}_B$, respectively.
The prime in ${\bf \nabla}'$ denotes that this nabla operator acts on ${\bf x}'$, hence ${\bf \nabla}'=(\partial_{x'},\partial_{y'},\partial_{z'})$ (in 3D) 
or ${\bf \nabla}'=(\partial_{x'},\partial_{z'})$ (in 2D).
The symbol $*_t$ stands for a temporal convolution, defined as  $f(t)*_tg(t)=\int_{-\infty}^\infty f(t-t')g(t'){\rm d}t'$.
When one of the functions is reversed in time, we obtain $f(t)*_tg(-t)=\int_{-\infty}^\infty f(t-t')g(-t'){\rm d}t'=\int_{-\infty}^\infty f(t+t')g(t'){\rm d}t'$.
Hence, $f(t)*_tg(-t)$ stands for a temporal correlation of the functions $f(t)$ and $g(t)$.
Evaluating the divergence ${\bf \nabla}'\cdot{\bf a}({\bf x}',t)$, using wave equation (\ref{eq31xg}) (with ${\bf x}$ replaced by ${\bf x}'$ and $t_A=0$)
and the property $\{\partial_t^2f(t)\}*_tg(-t)=f(t)*_t\{\partial_t^2g(-t)\}$, yields
\begin{eqnarray}
{\bf \nabla}'\cdot{\bf a}({\bf x}',t)={\cal G}_x({\bf x}',{\bf x}_B,t,0)\delta({\bf x}'-{\bf x}_A)-\delta({\bf x}'-{\bf x}_B){\cal G}_x({\bf x}',{\bf x}_A,-t,0).
\end{eqnarray}
Next, we integrate this over a domain ${\cal V}$, enclosed by boundary ${\cal S}$,
with outward pointing normal vector ${\bf n}$ 
(for the 3D situation, ${\cal V}$ is a 3D volume and ${\cal S}$ a 2D closed  boundary; 
for the 2D situation, ${\cal V}$ is a 2D plane and ${\cal S}$ a 1D closed  curve  in the $(x,z)$-plane). 
The Green's sources at ${\bf x}_A$ and ${\bf x}_B$ are both situated in ${\cal V}$.
Using the theorem of Gauss, $\int_{\cal V}{\bf \nabla}'\cdot{\bf a}{\rm d}{\bf x'}=\oint_{\cal S}{\bf a}\cdot{\bf n}{\rm d}{\bf x}'$,
applying the reciprocity relation of equation (\ref{eq31shfrec}) to  the Green's functions with the source at ${\bf x}_B$, and replacing ${\bf x}_B$ by variable ${\bf x}$ yields
the following representation of the homogeneous Green's function
\begin{eqnarray}
{\cal G}_x^h({\bf x},{\bf x}_A,t,0)&=&
\oint_{\cal S}\frac{1}{\beta({\bf x}')}\Bigl[\{{\bf \nabla}'{\cal G}_x({\bf x},{\bf x}',t,0)\}*_t{\cal G}_x({\bf x}',{\bf x}_A,-t,0)\nonumber\\
&&-{\cal G}_x({\bf x},{\bf x}',t,0)*_t{\bf \nabla}'{\cal G}_x({\bf x}',{\bf x}_A,-t,0)\Bigr]\cdot{\bf n}{\rm d}{\bf x}'.\label{eq1616}
\end{eqnarray}
This expression is the time-domain equivalent of the frequency-domain homogeneous Green's function representation, introduced for holographic imaging and 
inverse scattering \cite{Porter70JOSA, Oristaglio89IP}.
In the following we show how the homogeneous Green's function representation of equation (\ref{eq1616}) forms a theoretical basis for 
time-reversed acoustics and for Green's function retrieval.

\subsection{Time-reversed acoustics in an inhomogeneous, time-invariant material}\label{sec3C}

\begin{figure}[t]
\centerline{\epsfxsize=9 cm \epsfbox{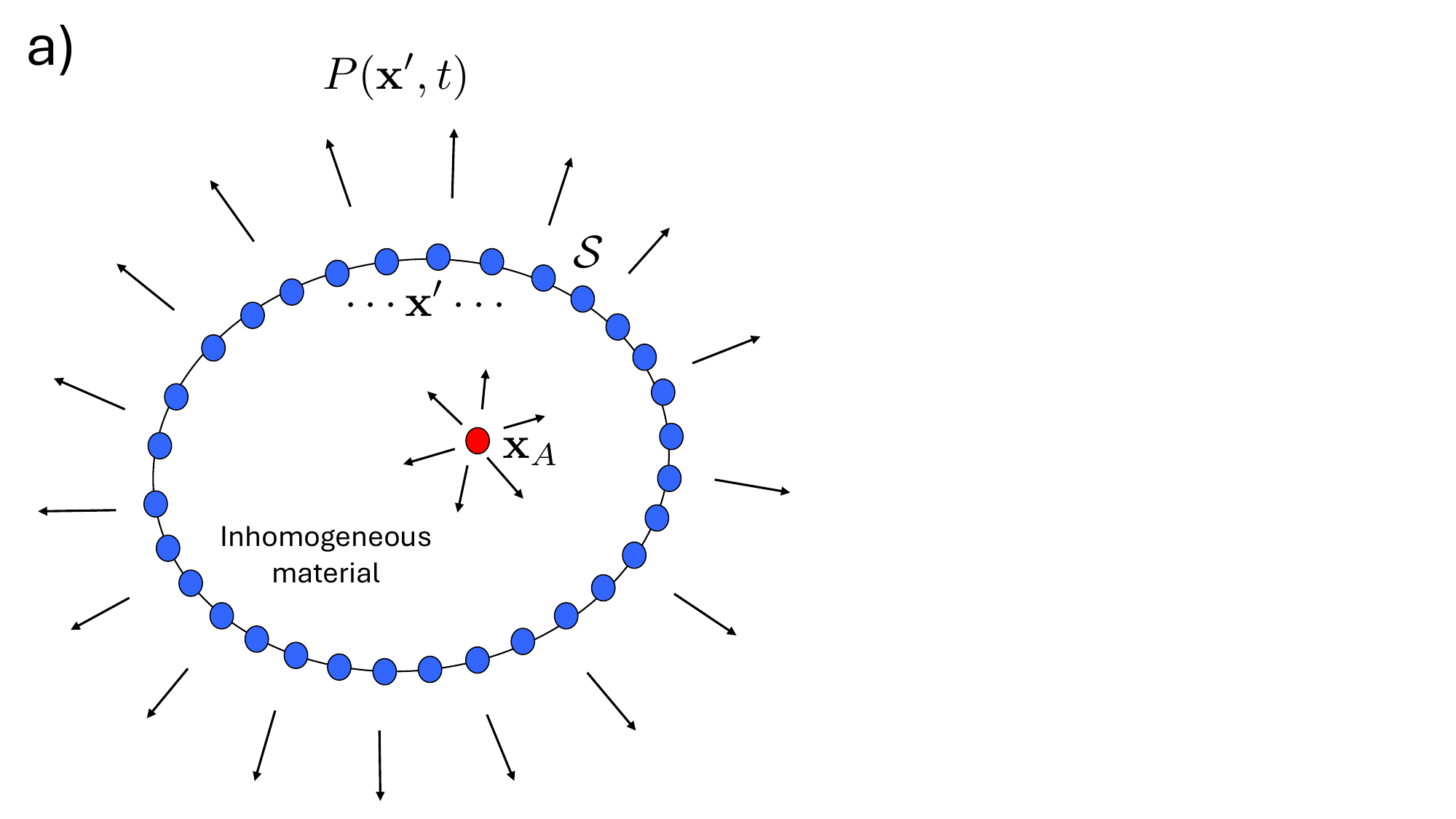}}
\vspace{0cm}
\centerline{\epsfxsize=9 cm \epsfbox{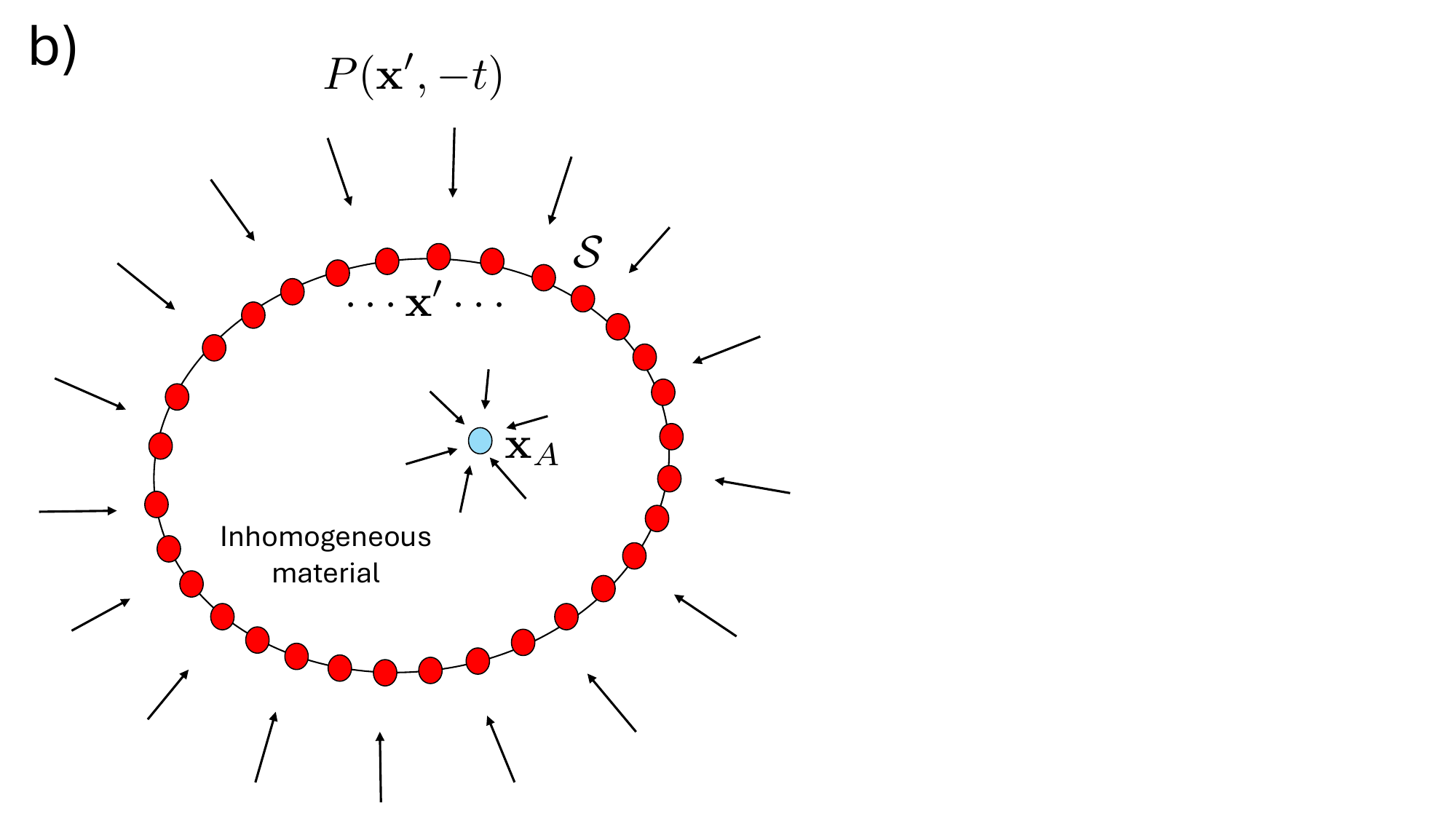}}
\vspace{0cm}
\centerline{\epsfxsize=9 cm \epsfbox{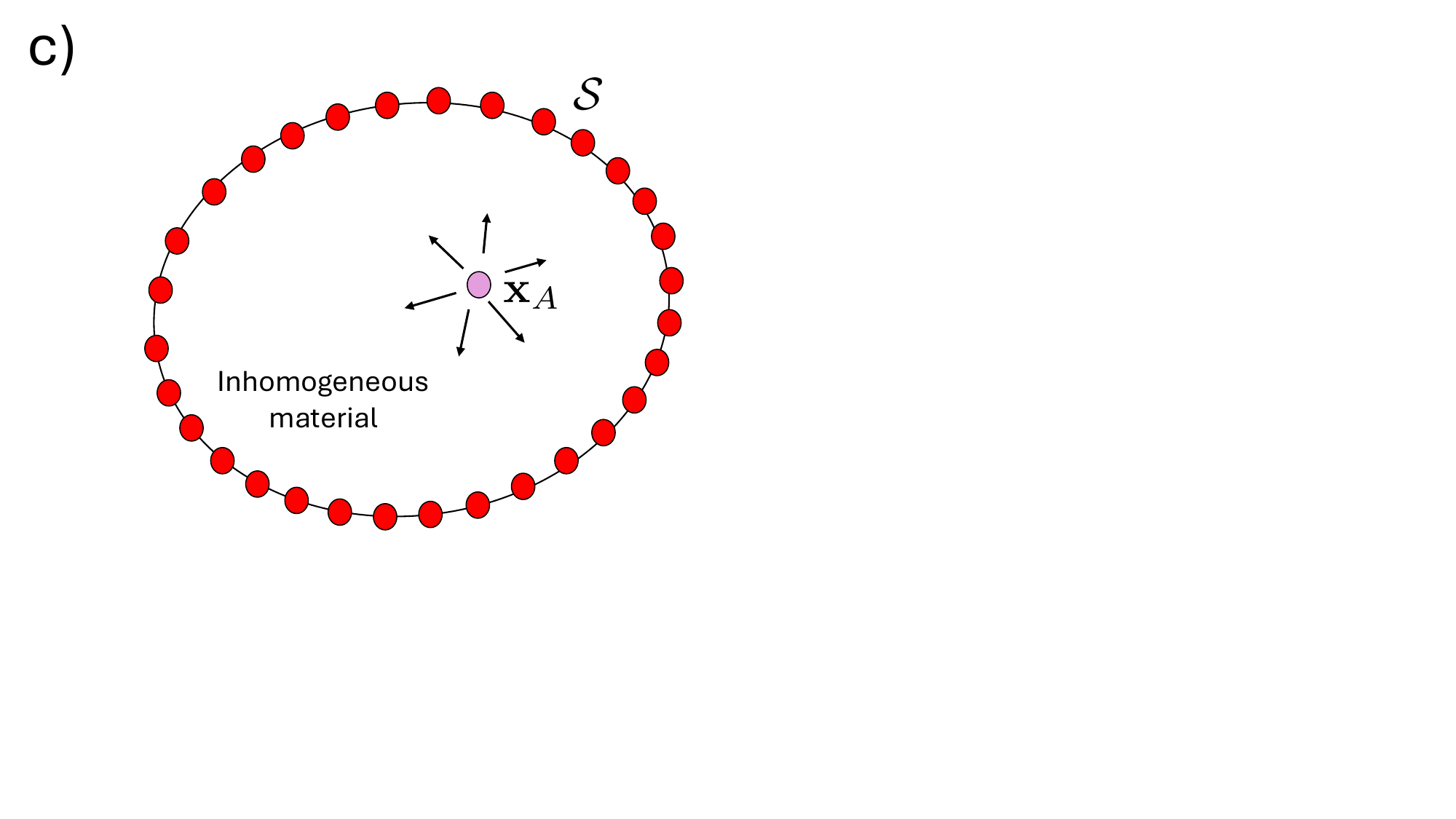}}
\vspace{-1.5cm}
\caption{Principle of time-reversed acoustics in an inhomogeneous, time-invariant material. (a) The response to a source at ${\bf x}_A$ and $t=0$ (indicated by the red dot), 
recorded by receivers at ${\bf x}'$.
(b) Time-reversed response, fed to sources at ${\bf x}'$. The wave field emitted by these sources focuses at $t=0$ at the original source position ${\bf x}_A$.
(c) After having focused, the wave field continues its propagation. The focused field at $t=0$  is  a virtual source (indicated by the pink dot) for the field for $t>0$.}\label{Figure1}
\end{figure}

We start by reviewing the principle of time-reversed acoustics in an inhomogeneous, time-invariant material, in an intuitive way.
Consider an impulsive point source at ${\bf x}_A$ and $t=0$ 
in an arbitrarily inhomogeneous, time-invariant material. The response to this source, $P({\bf x}',t)$, is recorded by receivers at ${\bf x}'$
on a closed boundary, see Figure \ref{Figure1}(a). In the acoustic situation, $P$ stands for the acoustic pressure.
Next, the responses at all ${\bf x}'$ are reversed in time, yielding $P({\bf x}',-t)$, and these time-reversed signals are fed to sources at the positions of the receivers,
see Figure \ref{Figure1}(b). 
The time-reversed wave field obeys the same wave equation as the original wave field. Hence, the time-reversed field emitted by the sources at the boundary
propagates back through the inhomogeneous material and  at $t=0$ it focuses at the original source position ${\bf x}_A$ (Figure \ref{Figure1}(b)).

In geophysics, this time-reversal principle is used to determine the parameters of earthquake sources \cite{McMechan82GJR, Gajewski2005GJI, Larmat2010PhysicsToday} 
and to perform seismic  imaging \cite{Hemon78GP, Whitmore83SEG, McMechan83GP, Zhou2018ESR}. 
Note that in geophysics the time-reversed field is not physically emitted into the earth, but instead the back-propagation is carried out numerically.

Fink and coworkers \cite{Fink92IEEE, Draeger99JASA, DeRosny2002PRL} developed the field of time-reversed acoustics for ultrasonic measurements,
in which piezoelectric transducers are initially used as receivers (as in Figure \ref{Figure1}(a)) and subsequently as sources (as in Figure \ref{Figure1}(b)),  
emitting the time-reversed responses physically into the material.
 
Note that once the field has focused at ${\bf x}_A$ and $t=0$, there is no sink to absorb the focused field, hence, the wave field continues its propagation \cite{Derode2003JASA}.
Hence, the focused field acts as a source which emits waves into the inhomogeneous material, see Figure \ref{Figure1}(c).
In other words, the focused field at $t=0$ is a virtual source for the field for $t>0$.

Let us now review the mathematics behind time-reversed acoustics, using the representation of equation (\ref{eq1616}) as the starting point  \cite{Fink2001IP}.
We start by assuming that the boundary ${\cal S}$ is sufficiently smooth and that the material at and outside ${\cal S}$ is homogeneous, 
with parameters $\alpha_0$, $\beta_0$ and $c_0$.
Then, in the high-frequency regime, the contribution of the two terms under the integral in equation (\ref{eq1616}) are identical (but with opposite signs) 
\cite{Wapenaar2005JASA}. Hence, we obtain
\begin{eqnarray}
{\cal G}_x^h({\bf x},{\bf x}_A,t,0)&=&
\frac{2}{\beta_0}\oint_{\cal S}\{{\bf \nabla}'{\cal G}_x({\bf x},{\bf x}',t,0)\cdot{\bf n}\}*_t{\cal G}_x({\bf x}',{\bf x}_A,-t,0){\rm d}{\bf x}'.\label{eq1515}
\end{eqnarray}
Inside  ${\cal S}$, we define a source function $s({\bf x},t)=\delta({\bf x}-{\bf x}_A)s_0(t)$, which is a point source at ${\bf x}={\bf x}_A$ with source wavelet $s_0(t)$.
From equations (\ref{eq31x}) and (\ref{eq31xg}) (with $t_A=0$) it follows that the response to this source, observed
at ${\bf x}'$ on ${\cal S}$,  is given by  $P({\bf x}',t)={\cal G}_x({\bf x}',{\bf x}_A,t,0)*_ts_0(t)$ (Figure \ref{Figure1}(a)).
Convolving both sides of equation (\ref{eq1515}) with the time-reversal of the source wavelet and substituting equation (\ref{eqghom}) into the left-hand side yields
\begin{eqnarray}
&&\{{\cal G}_x({\bf x},{\bf x}_A,t,0)-{\cal G}_x({\bf x},{\bf x}_A,-t,0)\}*_ts_0(-t)=\nonumber\\
&&\hspace{2cm}\frac{2}{\beta_0}\oint_{\cal S}
\underbrace{\{{\bf \nabla}'{\cal G}_x({\bf x},{\bf x}',t,0)\cdot{\bf n}\}}_{\mbox{\footnotesize``propagator''}}*_t
\underbrace{P({\bf x}',-t)}_{\mbox{\footnotesize ``source field''}}{\rm d}{\bf x}'.\label{eq1717}
\end{eqnarray}
The right-hand side quantifies the propagation of the time-reversed field $P({\bf x}',-t)$ from 
sources at ${\bf x}'$ on ${\cal S}$ to any point ${\bf x}$ inside ${\cal S}$.
The second term on the left-hand side, ${\cal G}_x({\bf x},{\bf x}_A,-t,0)*_ts_0(-t)$, is the field converging to ${\bf x}_A$ (Figure \ref{Figure1}(b)). 
The first term on the left-hand side, ${\cal G}_x({\bf x},{\bf x}_A,t,0)*_ts_0(-t)$, is the field diverging from ${\bf x}_A$ (Figure \ref{Figure1}(c)). 

In most practical situations, the object of investigation cannot be accessed from a closed boundary. 
Replacing the closed boundary integral in equation (\ref{eq1717}) by an open boundary integral
necessarily leads to approximations, which can be partly remedied by iterative methods.  
A further discussion is beyond the scope of this paper; we refer to the extensive literature 
on iterative time-reversed acoustics (for example references \cite{Prada95JASA, Montaldo2005IEEE, Borcea2007IP}) 
and on the Marchenko method (for example references \cite{Rose2002IP, Broggini2012EJP, Wapenaar2013PRL}). 

\subsection{Green's function retrieval in an inhomogeneous, time-invariant material}\label{sec3D}

\begin{figure}[t]
\centerline{\epsfxsize=9 cm \epsfbox{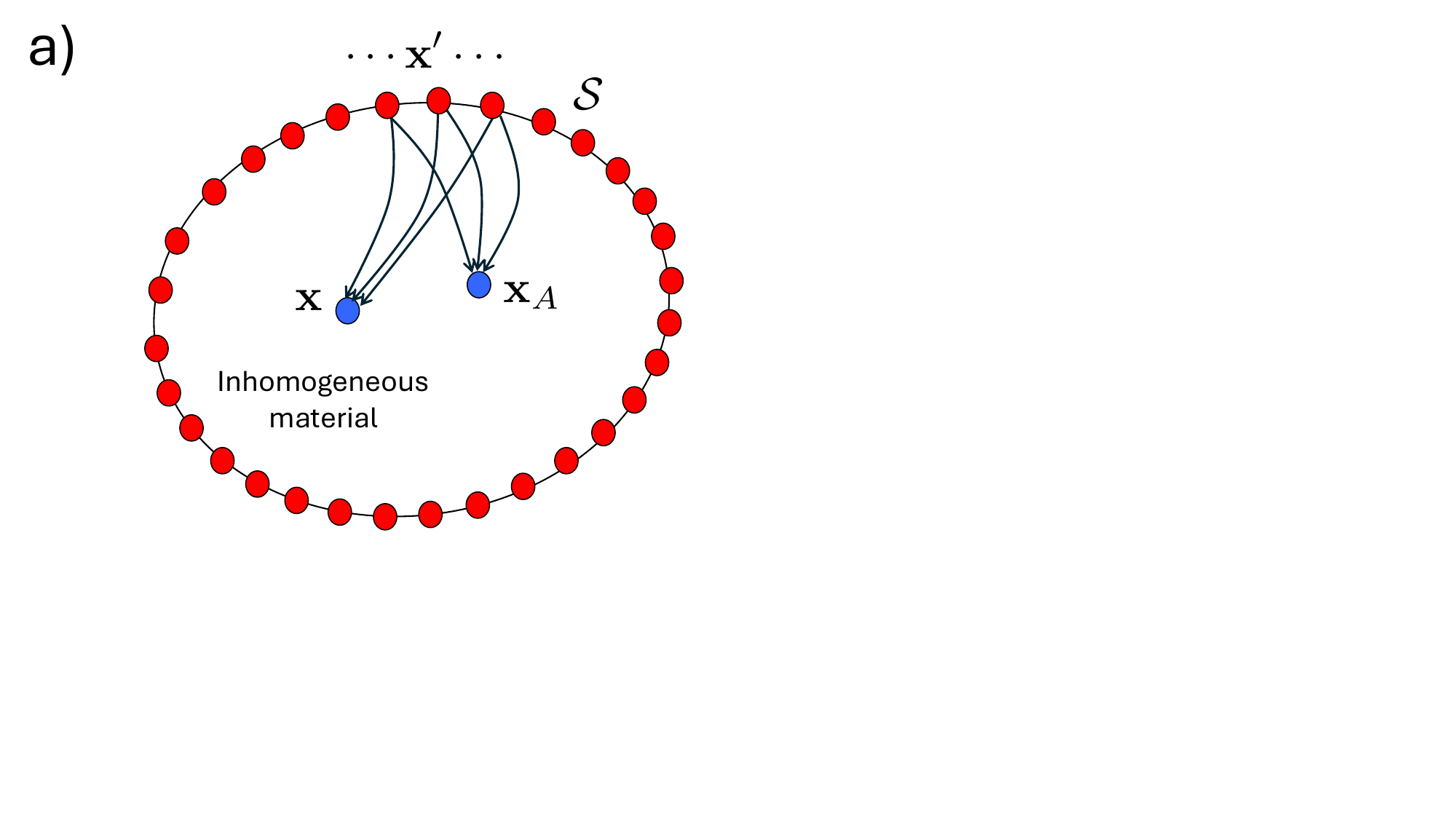}}
\vspace{-1.5cm}
\centerline{\epsfxsize=9 cm \epsfbox{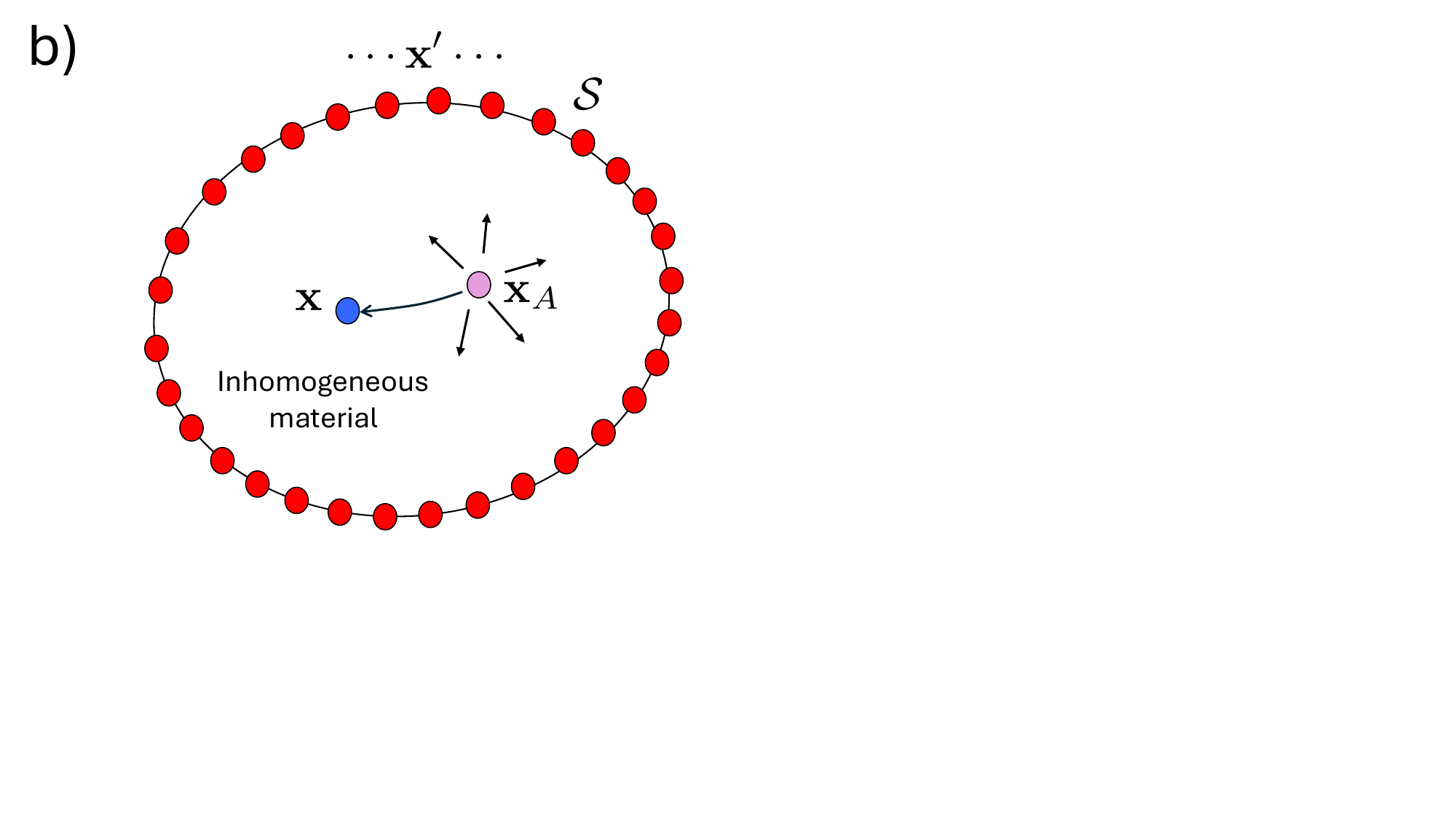}}
\vspace{-1.5cm}
\caption{Principle of Green's function retrieval in an inhomogeneous, time-invariant material. (a) Responses to impulsive sources at ${\bf x}'$ and $t=0$, 
recorded by receivers at ${\bf x}_A$ and ${\bf x}$.
(b) Response to a virtual source at ${\bf x}_A$ (indicated by the pink dot), recorded by a receiver at ${\bf x}$, obtained from the temporal crosscorrelation of the responses in (a), integrated over all
sources at ${\bf x}'$ on ${\cal S}$.}\label{Figure2}
\end{figure}

It has been shown by many authors that the temporal crosscorrelation of passive noise measurements at two receivers in an inhomogeneous, time-invariant material 
converges approximately to the response at one of these receivers as if there were a source at the position of the other (i.e., the Green's function between the two receivers)
\cite{Weaver2001PRL, Campillo2003Science, Wapenaar2003GEO, Snieder2004PRE, Malcolm2004PRE, Roux2004JASA2, Haney2009GRL, Sabra2007APL}.
We review the mathematics behind this principle of Green's function retrieval, using equation (\ref{eq1515}) as the starting point \cite{Wapenaar2003GEO, Wapenaar2005JASA}.
Using reciprocity relation (\ref{eq31shfrec}) and substituting equation (\ref{eqghom}) into the left-hand side yields
\begin{eqnarray}
&&{\cal G}_x({\bf x},{\bf x}_A,t,0)-{\cal G}_x({\bf x},{\bf x}_A,-t,0)=\nonumber\\
&&\hspace{2cm}\frac{2}{\beta_0}\oint_{\cal S}\{{\bf \nabla}'{\cal G}_x({\bf x},{\bf x}',t,0)\cdot{\bf n}\}*_t{\cal G}_x({\bf x}_A,{\bf x}',-t,0){\rm d}{\bf x}'.\label{eq1919}
\end{eqnarray}
The Green's function ${\cal G}_x({\bf x}_A,{\bf x}',t,0)$ on the right-hand side represents the response  to an impulsive monopole source at ${\bf x}'$ on ${\cal S}$ and $t=0$, 
recorded by a receiver at ${\bf x}_A$. 
Similarly, ${\bf \nabla}'{\cal G}_x({\bf x},{\bf x}',t,0)\cdot{\bf n}$ (with nabla operator ${\bf \nabla}'$ acting on the source position ${\bf x}'$) 
represents the response  to an impulsive dipole source at ${\bf x}'$ on ${\cal S}$ and $t=0$, 
recorded by a receiver at ${\bf x}$. Both responses are illustrated in Figure \ref{Figure2}(a). The entire right-hand side quantifies the temporal crosscorrelation of these
responses at ${\bf x}_A$ and ${\bf x}$, integrated over all sources at ${\bf x}'$ on ${\cal S}$.
The left-hand side consists of the Green's function ${\cal G}_x({\bf x},{\bf x}_A,t,0)$ (minus its time-reversed version).
This is the response to an impulsive virtual source at ${\bf x}_A$ and $t=0$, recorded by a physical receiver at ${\bf x}$, see Figure \ref{Figure2}(b).

When only monopole sources are available on ${\cal S}$, we need to approximate the dipole responses on the right-hand side of equation (\ref{eq1919}) by monopole responses.
We already assumed that the material at and outside ${\cal S}$ is homogeneous, with parameters $\alpha_0$, $\beta_0$ and $c_0$.
If we assume in addition that ${\cal S}$ is a sphere with very large radius, we may use the following approximation on ${\cal S}$ \cite{Wapenaar2005JASA}
\begin{eqnarray}
{\bf \nabla}'{\cal G}_x({\bf x},{\bf x}',t,0)\cdot{\bf n}&\approx&-\frac{1}{c_0}\partial_t{\cal G}_x({\bf x},{\bf x}',t,0).
\end{eqnarray}
Substituting this into equation (\ref{eq1919}), using $\{\partial_tf(t)\}*_tg(t)=\partial_t\{f(t)*_tg(t)\}$, yields
\begin{eqnarray}
&&{\cal G}_x({\bf x},{\bf x}_A,t,0)-{\cal G}_x({\bf x},{\bf x}_A,-t,0)\approx\nonumber\\
&&\hspace{3cm}
-\frac{2}{\beta_0c_0}\partial_t\oint_{\cal S}{\cal G}_x({\bf x},{\bf x}',t,0)*_t{\cal G}_x({\bf x}_A,{\bf x}',-t,0){\rm d}{\bf x}'.\label{eq22}
\end{eqnarray}
Now the right-hand side quantifies the time-derivative of the crosscorrelation of monopole responses, integrated over the sources.
An alternative form, without the time-derivative, reads \cite{Wapenaar2006GEO}
\begin{eqnarray}
&&G_x({\bf x},{\bf x}_A,t,0)+G_x({\bf x},{\bf x}_A,-t,0)\approx\nonumber\\
&&\hspace{3cm}
\frac{2}{\beta_0c_0}\oint_{\cal S}G_x({\bf x},{\bf x}',t,0)*_tG_x({\bf x}_A,{\bf x}',-t,0){\rm d}{\bf x}',\label{eq22alt}
\end{eqnarray}
where the Green's function $G_x({\bf x},{\bf x}_A,t,0)$ is the response to a source $a({\bf x},t)=\delta({\bf x}-{\bf x}_A)\delta(t)$, ${\bf b}({\bf x},t)={\bf 0}$.
Since $s({\bf x},t)=\partial_ta({\bf x},t)$ (equation \ref{eqsourcex}) and the material is time-invariant, 
we have $G_x({\bf x},{\bf x}_A,t,0)=\partial_t{\cal G}_x({\bf x},{\bf x}_A,t,0)$.
Expressions (\ref{eq22}) and (\ref{eq22alt}) are useful when the responses to all sources at ${\bf x}'$ on ${\cal S}$ are individually available,
such as in controlled-source seismic interferometry \citep{Bakulin2006GEO, Schuster2009Book, Neut2011GEO}. 
Of course approximations are introduced when the sources are not available on a closed boundary, similar as in
time-reversed acoustics, as noted in section \ref{sec3C}.

For the derivation of Green's function retrieval from ambient noise,
consider the situation in which the sources at all ${\bf x}'$ on ${\cal S}$ are simultaneously acting noise sources $N({\bf x}',t)$. 
In that case, the fields at ${\bf x}_A$ and ${\bf x}$ are given by
$P({\bf x}_A,t)=\oint_{{\cal S}}{\cal G}_x({\bf x}_A,{\bf x}',t,0)*_tN({\bf x}',t){\rm d}{\bf x}'$ and 
$P({\bf x},t)=\oint_{{\cal S}}{\cal G}_x({\bf x},{\bf x}'',t,0)*_tN({\bf x}'',t){\rm d}{\bf x}''$,
respectively. We assume that the noise sources are mutually uncorrelated,
according to $\langle N({\bf x}'',t)*_tN({\bf x}',-t)\rangle=\delta({\bf x}'-{\bf x}'')C(t)$. 
Here $C(t)$ is the temporal autocorrelation of the noise (which is assumed to be the same for all sources)
and $\langle\cdot\rangle$ stands for ensemble averaging, which in practice is replaced by averaging over very long time. 
Further,  $\delta({\bf x}'-{\bf x}'')$ is a delta function defined in ${\cal S}$.
The crosscorrelation of the fields at ${\bf x}_A$ and ${\bf x}$ thus gives \cite{Wapenaar2006GEO}
\begin{eqnarray}\label{eq35}
\hspace{-.0cm}\langle P({\bf x},t)*_tP({\bf x}_A,-t)\rangle
&=&\oint_{{\cal S}}{\cal G}_x({\bf x},{\bf x}',t,0)*_t{\cal G}_x({\bf x}_A,{\bf x}',-t,0)*_tC(t){\rm d}{\bf x}'.
\end{eqnarray}
Convolving both sides of equation (\ref{eq22}) with the autocorrelation $C(t)$,
using $\{\partial_tf(t)\}*_tg(t)=\partial_t\{f(t)*_tg(t)\}$ and  equation (\ref{eq35}) to simplify the right-hand side, we obtain
\begin{equation}\label{eq3SSIb}
\{{\cal G}_x({\bf x},{\bf x}_A,t,0)-{\cal G}_x({\bf x},{\bf x}_A,-t,0)\}*_tC(t)\approx -\frac{2}{\beta_0 c_0}\partial_t\langle P({\bf x},t)*_tP({\bf x}_A,-t)\rangle.
\end{equation}
This is the basic expression for Green's function retrieval from noise, also known as ambient-noise interferometry 
\citep{Weaver2001PRL, Derode2003JASA, Wapenaar2003GEO, Campillo2003Science, Snieder2004PRE, Wapenaar2006GEO}. 
It states that the time-derivative of the crosscorrelation of two noise recordings at receiver positions ${\bf x}_A$ and ${\bf x}$, 
averaged over  long enough  time, converges to the 
Green's function between ${\bf x}_A$ and ${\bf x}$ (minus its time-reversed version), convolved with the temporal autocorrelation of the noise.
Hence, the receiver at ${\bf x}_A$ is turned into a virtual source.

In most practical situations the sources are not distributed along a closed boundary, the noise sources may be mutually correlated, 
the noise spectra may not be the same for all sources, 
the wave fields may not obey a simple scalar wave equation, etc.
Hence, the expressions derived in this section are at best approximations for real situations. Much research has been done to overcome these limitations, of which we mention
iterative correlation \cite{Stehly2008JGR}, multidimensional deconvolution \cite{Wapenaar2010JASA, Neut2011GEO}, directional balancing \cite{Curtis2010GEO}
and generalized interferometry \cite{Fichtner2017GJI}. A further discussion is beyond the scope of this paper.

\section{Time-reversed acoustics and Green's function retrieval in homogeneous, time-variant materials}\label{sec4} 

We discuss time-reversed acoustics and Green's function retrieval in homogeneous, time-variant materials, and their mutual relation.
The setup is analogous to that in section \ref{sec3} for inhomogeneous, time-invariant materials, which facilitates the comparison of the methodologies for both classes of material.
In section \ref{sec4A} we discuss the wave equation for a homogeneous, time-variant material, as a special case of the equations in section \ref{sec2}.
In section \ref{sec4B} we discuss the Green's function of a homogeneous, time-variant material.
In section \ref{sec4C} we discuss the propagator matrix for a homogeneous, time-variant material.
In section \ref{sec4H} we use this propagator matrix as the basis for deriving  a representation of the homogeneous Green's function of a homogeneous, time-variant material.
In sections \ref{sec4D} and \ref{sec4E} we show how this homogeneous Green's function representation leads to 
time-reversed acoustics and Green's function retrieval, respectively, in homogeneous, time-variant materials.

\subsection{Wave equation for a homogeneous, time-variant material}\label{sec4A}
We consider a homogeneous, arbitrarily time-variant material, with parameters  $\alpha(t)$ and $\beta(t)$.
Substituting constitutive relations (\ref{eq3}) and (\ref{eq4}) into equations (\ref{eq1}) and (\ref{eq2}), 
using the fact that $\alpha(t)$ and $\beta(t)$ are space-invariant, we obtain
\begin{eqnarray}
\partial_t U + \frac{1}{\beta}{\bf \nabla}\cdot {\bf V} &=& a,\label{eq49}\\
\partial_t {\bf V} + \frac{1}{\alpha}{\bf \nabla} U &=& {\bf b}.\label{eq50}
\end{eqnarray}
For acoustic waves these expressions become $-\partial_t\Theta + \frac{1}{\rho}{\bf \nabla}\cdot {\bf m}=q$ and $\partial_t{\bf m}-\frac{1}{\kappa}{\bf \nabla}\Theta={\bf f}$.
We continue with the general notation of equations (\ref{eq49}) and (\ref{eq50}), so that all that follows also holds for 2D SH, TE and TM waves.

The parameters $\alpha(t)$ and $\beta(t)$ in equations (\ref{eq49}) and (\ref{eq50}) may vary continuously with time.
For a material with piecewise continuous parameters, these equations are supplemented with boundary conditions.
At a time boundary, where the parameters undergo a finite jump, the boundary conditions
state that $U$ and ${\bf V}$ are continuous over that boundary.

By eliminating ${\bf V}$ from equations (\ref{eq49}) and (\ref{eq50}) we obtain the second order wave equation
\begin{eqnarray}
\bigl(\partial_t\beta\partial_t - \beta c^2{\bf \nabla}^2\bigr)U({\bf x},t) = s({\bf x},t),\label{eq31a}
\end{eqnarray}
with  $c(t)$ being the time-variant propagation velocity, defined again as $c=\frac{1}{\sqrt{\alpha\beta}}$, and $s({\bf x},t)$ the source distribution, defined as
\begin{eqnarray}
s=\partial_t(\beta a) - {\bf \nabla}\cdot {\bf b}.\label{eqsource}
\end{eqnarray}

\subsection{Green's function of a homogeneous, time-variant material}\label{sec4B}

We introduce the Green's function ${\cal G}_t({\bf x},{\bf x}_A,t,t_A)$ of a homogeneous, time-variant 
material as the response to an impulsive point source at ${\bf x}={\bf x}_A$ at time instant $t=t_A$,
observed at position ${\bf x}$ and time $t$. Hence, it obeys wave
equation (\ref{eq31a}), with the source term $s({\bf x},t)$ on the right-hand side replaced by an impulsive point source, according to
\begin{eqnarray}
\bigl(\partial_t\beta\partial_t - \beta c^2{\bf \nabla}^2\bigr){\cal G}_t({\bf x},{\bf x}_A,t,t_A)= \delta({\bf x}-{\bf x}_A)\delta(t-t_A).\label{eq31agbl}
\end{eqnarray}
The subscript $t$ in ${\cal G}_t$ denotes that this is the Green's function of a time-variant material.
The Green's function obeys the causality condition
\begin{eqnarray}
{\cal G}_t({\bf x},{\bf x}_A,t,t_A)=0 \quad \mbox{for}\quad t<t_A.\label{eq31con}
\end{eqnarray}
Assuming that the material is time-invariant after an arbitrarily large but finite time, this causality condition implies that ${\cal G}_t({\bf x},{\bf x}_A,t,t_A)$ is
outward propagating for $|{\bf x}-{\bf x}_A|\to\infty$.
Analytical expressions of ${\cal G}_t({\bf x},{\bf x}_A,t,t_A)$ for the special case of a homogeneous, time-invariant material, are given in Appendix \ref{AppA1}.
Here we continue with expressions for a  time-variant material.

Since the material is homogeneous, the Green's function is space-shift invariant, hence
\begin{eqnarray}
{\cal G}_t({\bf x},{\bf x}_A,t,t_A)={\cal G}_t({\bf x}-{\bf x}_A,{\bf 0},t,t_A).\label{eq31shf}
\end{eqnarray}

We introduce an acausal Green's function ${\cal G}_t^a({\bf x},{\bf 0},t,t_A)$, 
obeying the same wave equation as the causal Green's function (equation (\ref{eq31agbl}), with ${\bf x}_A={\bf 0}$), 
but with the acausality condition
\begin{eqnarray}
{\cal G}_t^a({\bf x},{\bf 0},t,t_A)=0 \quad \mbox{for}\quad t>t_A.\label{eq31conac}
\end{eqnarray}
Assuming that the material is time-invariant before an arbitrarily large but finite negative time, this acausality condition implies that
${\cal G}_t^a({\bf x},{\bf 0},t,t_A)$ is inward propagating for $|{\bf x}|\to\infty$. Subsequently, it propagates 
through the time-variant material, converges to a sink at ${\bf x}={\bf 0}$ and $t=t_A$, after which it disappears.
Note, however, that since the material is time-variant, in general this acausal Green's function is not the time-reversal of the causal Green's function,
i.e., ${\cal G}_t^a({\bf x},{\bf 0},t,0)\ne{\cal G}_t({\bf x},{\bf 0},-t,0)$.

The causal and acausal Green's functions are mutually related via the reciprocity relation \cite{Wapenaar2025RS}
\begin{eqnarray}
{\cal G}_t({\bf x},{\bf 0},t_A,t_B)={\cal G}_t^a({\bf x},{\bf 0},t_B,t_A).\label{eq1022sqshbl}
\end{eqnarray}
Hence, in a homogeneous, time-variant material, the causal response to a source at $t_B$, observed at $t_A$, is equal to the acausal response observed at $t_B$, 
absorbed by a sink at $t_A$. This is a meaningful relation for $t_A>t_B$. For $t_A<t_B$, equation (\ref{eq1022sqshbl}) reduces to the trivial relation $0=0$,
on account of equations (\ref{eq31con}) and (\ref{eq31conac}).

We define the homogeneous Green's function of a homogeneous, time-variant material, as the difference of the causal and acausal Green's functions, according to
\begin{eqnarray}
{\cal G}_t^h({\bf x},{\bf 0},t,t_A)={\cal G}_t({\bf x},{\bf 0},t,t_A)-{\cal G}_t^a({\bf x},{\bf 0},t,t_A).\label{eq1022sqshblhom}
\end{eqnarray}
Both terms on the right-hand side obey the same wave equation, with the same singularity $\delta({\bf x})\delta(t-t_A)$.
The singularities cancel when we subtract the equations for the causal and acausal  Green's functions. Hence, the homogeneous Green's function
obeys the following equation
\begin{eqnarray}
\bigl(\partial_t\beta\partial_t - \beta c^2{\bf \nabla}^2\bigr){\cal G}_t^h({\bf x},{\bf 0},t,t_A)= 0.\label{eq31agblhom}
\end{eqnarray}
Unlike the Green's function ${\cal G}_t({\bf x},{\bf 0},t,t_A)$ and its acausal counterpart  ${\cal G}_t^a({\bf x},{\bf 0},t,t_A)$, 
the homogeneous Green's function ${\cal G}_t^h({\bf x},{\bf 0},t,t_A)$ is not singular for ${\bf x}\to{\bf 0}$ and $t\to t_A$.

From equations (\ref{eq1022sqshbl}) and (\ref{eq1022sqshblhom}) it follows that the homogeneous Green's function obeys the following reciprocity relation
\begin{eqnarray}
{\cal G}_t^h({\bf x},{\bf 0},t_A,t_B)=-{\cal G}_t^h({\bf x},{\bf 0},t_B,t_A).\label{eq1022sqshblkkk}
\end{eqnarray}

In the next section we discuss the propagator matrix for a homogeneous, time-variant material. The elements of this matrix can be expressed in terms of the 
homogeneous Green's function ${\cal G}_t^h({\bf x},{\bf 0},t,t_A)$. The propagator matrix formalism will be used in section \ref{sec4H} as a basis for deriving
a representation of the  homogeneous Green's function for a homogeneous, time-variant material 
(as the counterpart of the representation of equation (\ref{eq1616}) for an inhomogeneous, time-invariant material).
 
\subsection{Propagator matrix of a homogeneous, time-variant material}\label{sec4C}

We reorganize equations (\ref{eq49}) and (\ref{eq50}) into a matrix-vector wave equation for a homogeneous, time-variant material, 
as the counterpart of a matrix-vector wave equation for an inhomogeneous, time-invariant material 
\cite{Corones75JMAA, Kosloff83GEO, Fishman84JMP, Ursin83GEO, Wapenaar86GP2, Loseth2007GJI}.
This equation will form the basis for discussing the propagator matrix of a homogeneous, time-variant material.

We reduce the number of field quantities in equations (\ref{eq49}) and (\ref{eq50}) by defining a scalar wave field $V_{\rm d}({\bf x},t)$ as
\begin{eqnarray}
V_{\rm d}&=&{\bf \nabla}\cdot{\bf V},\label{eq457b}
\end{eqnarray}
where subscript ${\rm d}$ refers to ``divergence''.  Using this in equations (\ref{eq49}) and (\ref{eq50}), we obtain
\begin{eqnarray}
\partial_t U + \frac{1}{\beta}V_{\rm d}&=& a,\label{eq49b}\\
\partial_t V_{\rm d}+ \frac{1}{\alpha}{\bf \nabla}^2 U &=& b_{\rm d},\label{eq50b}
\end{eqnarray}
with
\begin{eqnarray}
b_{\rm d}&=&{\bf \nabla}\cdot{\bf b}.\label{eq458b}
\end{eqnarray}
Equations (\ref{eq49b}) and (\ref{eq50b})  can be reorganized into a matrix-vector wave equation, according to 
\begin{eqnarray}
\partial_t{\bf q}={\bf A}{\bf q}+{\bf d},\label{eq67blue}
\end{eqnarray}
with 
\begin{eqnarray}
{\bf q}=\begin{pmatrix}   U\\   V_{\rm d}\end{pmatrix},\quad
{\bf A}=\begin{pmatrix} 0 & -\frac{1}{\beta}\\-\frac{1}{\alpha}{\bf \nabla}^2& 0 \end{pmatrix},\quad
{\bf d}=\begin{pmatrix}   a\\   b_{\rm d}\end{pmatrix},\label{eq68blue}
\end{eqnarray}
where ${\bf q}({\bf x},t)$ is a wave field vector, ${\bf A}(t)$ an operator matrix and  ${\bf d}({\bf x},t)$ a source vector.
The parameters $\alpha(t)$ and $\beta(t)$ in operator matrix ${\bf A}(t)$ may vary continuously with time.
For a material with piecewise continuous parameters, equation (\ref{eq67blue}) is supplemented with boundary conditions:
at each time boundary, the wave field vector ${\bf q}({\bf x},t)$ is continuous over that boundary.

We introduce the $2\times 2$ propagator matrix ${\bf W}({\bf x},t,t_A)$ as the solution of equation (\ref{eq67blue}) for the source-free situation 
\cite{Salem2015arXiv, Torrent2018PRB,  Wapenaar2025RS}, hence
\begin{eqnarray}
\partial_t{\bf W}={\bf A}{\bf W}.\label{eq31}
\end{eqnarray}
For $t=t_A$ it obeys the initial condition
\begin{eqnarray}
{\bf W}({\bf x},t_A,t_A)={\bf I}\delta({\bf x}),\label{eq33}
\end{eqnarray}
where ${\bf I}$ is the identity matrix. 
The propagator matrix ${\bf W}({\bf x},t,t_A)$ forms the counterpart of the propagator matrix for an inhomogeneous, time-invariant material
 \cite{Thomson50JAP, Haskell53BSSA, Gilbert66GEO, Woodhouse74GJR, Kennett79GJRAS}.
In the literature on time-variant materials, this matrix is also known as the transfer matrix  \cite{Torrent2018PRB, Salem2015arXiv, Pacheco2020NP}.
From equations (\ref{eq67blue}), (\ref{eq31}) and (\ref{eq33}), assuming $\alpha(t)$ and $\beta(t)$ vary continuously and  
there are no sources between $t_A$ and $t$, it follows that ${\bf q}({\bf x},t)$ can be expressed as
\begin{eqnarray}
{\bf q}({\bf x},t)={\bf W}({\bf x},t,t_A)*_x{\bf q}({\bf x},t_A).\label{eqprop}
\end{eqnarray}
Here $t$ may be larger or smaller than, or equal to $t_A$.
The symbol $*_x$ stands for a 3D or 2D spatial convolution, according to $f({\bf x})*_xg({\bf x})=\int_{\mathbb{R}^m} f({\bf x}-{\bf x}')g({\bf x}'){\rm d}{\bf x}'$,
with $m=3$ for the 3D situation and $m=2$ for the 2D situation; $\mathbb{R}$ denotes the set of real numbers.

Let $t_1\dots t_n\dots t_{N-1}$ be time instants where the parameters $\alpha(t)$ and $\beta(t)$ may be discontinuous. Then, by applying equation (\ref{eqprop}) recursively
(starting with $t_A=t_0$), we obtain
\begin{eqnarray}
{\bf W}({\bf x},t_N,t_0)={\bf W}({\bf x},t_N,t_{N-1})*_x\cdots*_x{\bf W}({\bf x},t_n,t_{n-1})*_x\cdots*_x{\bf W}({\bf x},t_1,t_0).\label{eqrecur}
\end{eqnarray}
This expression is useful for numerical modeling, in which case the $t_n$ are ordered from small to large (an example is given at the end of this section). 
However, this expression remains valid for arbitrarily ordered $t_n$.

We partition the $2\times 2$ matrix ${\bf W}({\bf x},t,t_A)$ as follows
\begin{eqnarray}
{\bf W}({\bf x},t,t_A)=
\begin{pmatrix} W^{U,U} &  W^{U,V} \\ & \\  W^{V,U} &  W^{V,V} \end{pmatrix}({\bf x},t,t_A).\label{eq34}
\end{eqnarray}
According to equation (11.27) in reference \cite{Wapenaar2025RS}, $W^{U,V}({\bf x},t,t_A)$ is related to the causal and acausal Green's functions via
\begin{eqnarray}
 W^{U,V}({\bf x},t,t_A)&=& {\cal G}_t^a({\bf x},{\bf 0},t,t_A)- {\cal G}_t({\bf x},{\bf 0},t,t_A)\nonumber\\
 &=&-{\cal G}_t^h({\bf x},{\bf 0},t,t_A).\label{eq35aa}
\end{eqnarray}
According to equations (11.18), (11.20) and (11.21) in reference \cite{Wapenaar2025RS}, the other components of the propagator matrix can be expressed in terms of 
$W^{U,V}({\bf x},t,t_A)$. Combining those relations with equation (\ref{eq35aa}) we obtain
\begin{eqnarray}
W^{V,V}({\bf x},t,t_A)&=&\beta(t)\partial_t {\cal G}_t^h({\bf x},{\bf 0},t,t_A),\label{eq39}\\
W^{U,U}({\bf x},t,t_A)&=&-\beta(t_A)\partial_{t_A} {\cal G}_t^h({\bf x},{\bf 0},t,t_A),\label{eq40}\\
W^{V,U}({\bf x},t,t_A)&=&\beta(t)\beta(t_A)\partial_t\partial_{t_A} {\cal G}_t^h({\bf x},{\bf 0},t,t_A).\label{eq41}
\end{eqnarray}
Equations (\ref{eq34}) -- (\ref{eq41}) show that the propagator matrix is expressed in terms of the 
Green's function and its derivatives. This property is restricted to homogeneous, time-variant materials
(i.e., it does not hold for the analogous case of inhomogeneous, time-invariant materials). 
This is explained as follows.
For inhomogeneous, time-invariant materials, the initial condition of the propagator matrix (equation (\ref{eq33})) 
is replaced by a boundary condition, but the causality condition of the Green's function (equation (\ref{eq31con})) remains
valid (i.e., it is not replaced by a boundary condition).
Therefore, since time and space are  interchangeable for the propagator matrix but not for the Green's function,
the relations between the propagator matrix and the  Green's function, as expressed by
equations (\ref{eq35aa}) -- (\ref{eq41}), have no counterpart for  inhomogeneous, time-invariant materials.
In section \ref{sec4H} we make use of the simple relations between the propagator matrix and the 
Green's function to derive a representation of the homogeneous Green's function from a representation of the propagator matrix.

\begin{figure}[t]
\vspace{-1.5cm}
\centerline{\hspace{4cm}\epsfxsize=13 cm \epsfbox{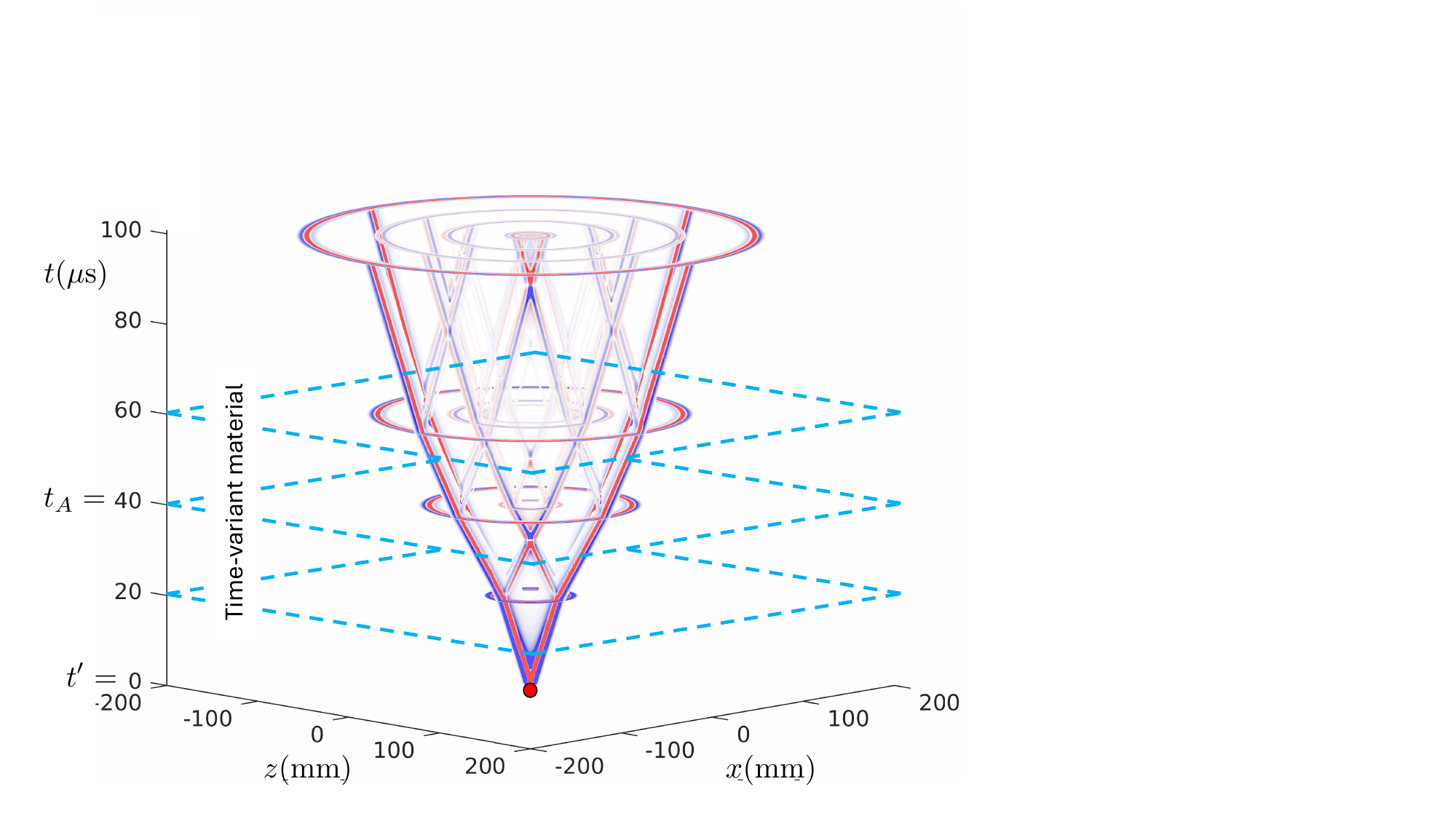}}
\caption{2D Green's function ${\cal G}_t({\bf x},{\bf 0},t,t'=0)$ (convolved with a spatial wavelet) in a piecewise constant, time-variant material.
The red dot indicates the source at ${\bf x}={\bf 0}$ and $t'=0$. The dashed blue planes indicate the time boundaries.
Movie available at https://www.keeswapenaar.nl/TimeMaterial/Green5.mp4}\label{Figure3}
\end{figure}

\begin{figure}[t]
\centerline{\hspace{7cm}\epsfxsize=16 cm \epsfbox{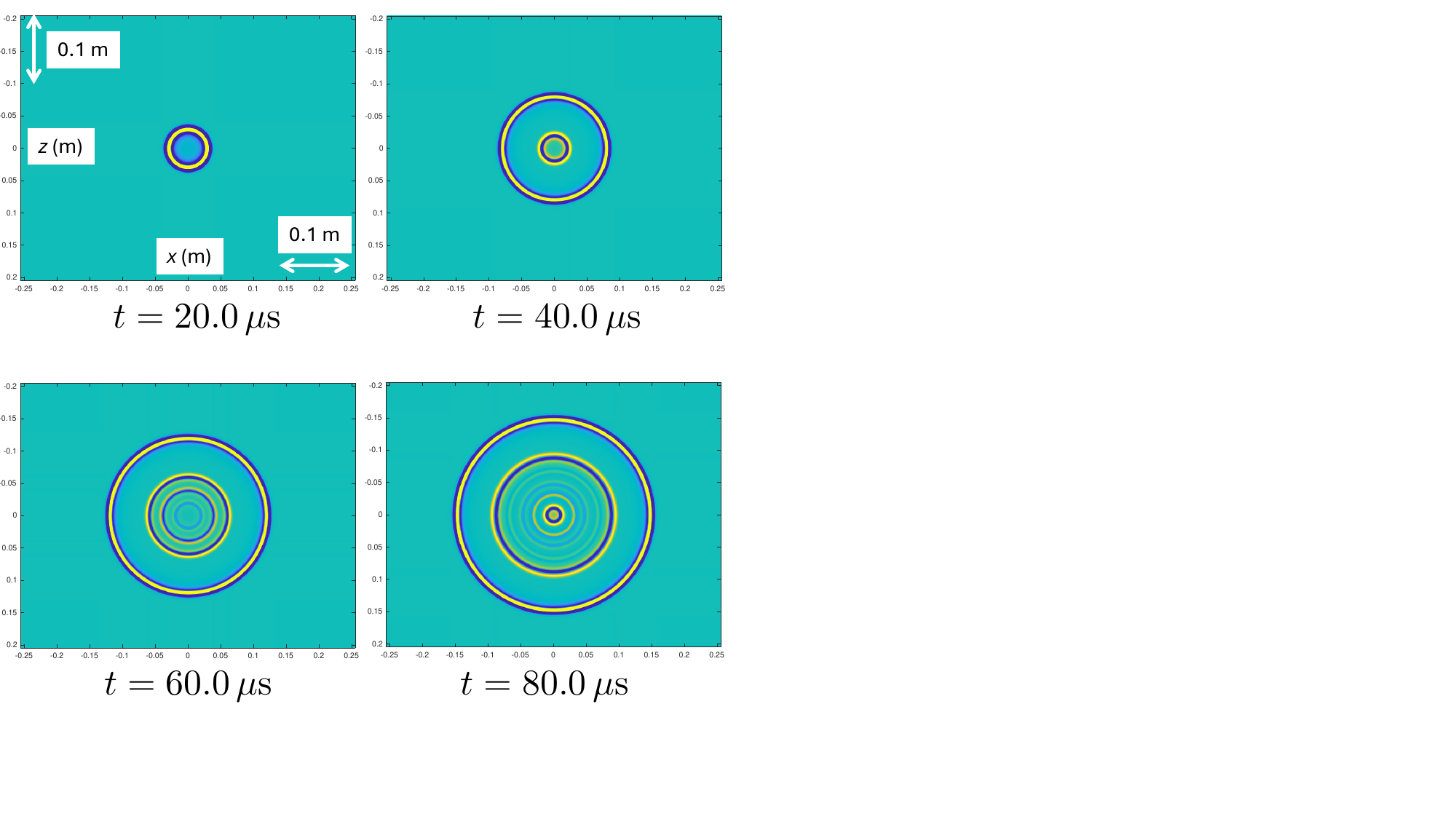}}
\vspace{-1.5cm}
\caption{Snapshots of the Green's function of Figure \ref{Figure3} for constant $t$ as a function of ${\bf x}=(x,z)$. }\label{Figure4}
\end{figure}

\begin{figure}[t]
\centerline{\hspace{5cm}\epsfxsize=15 cm \epsfbox{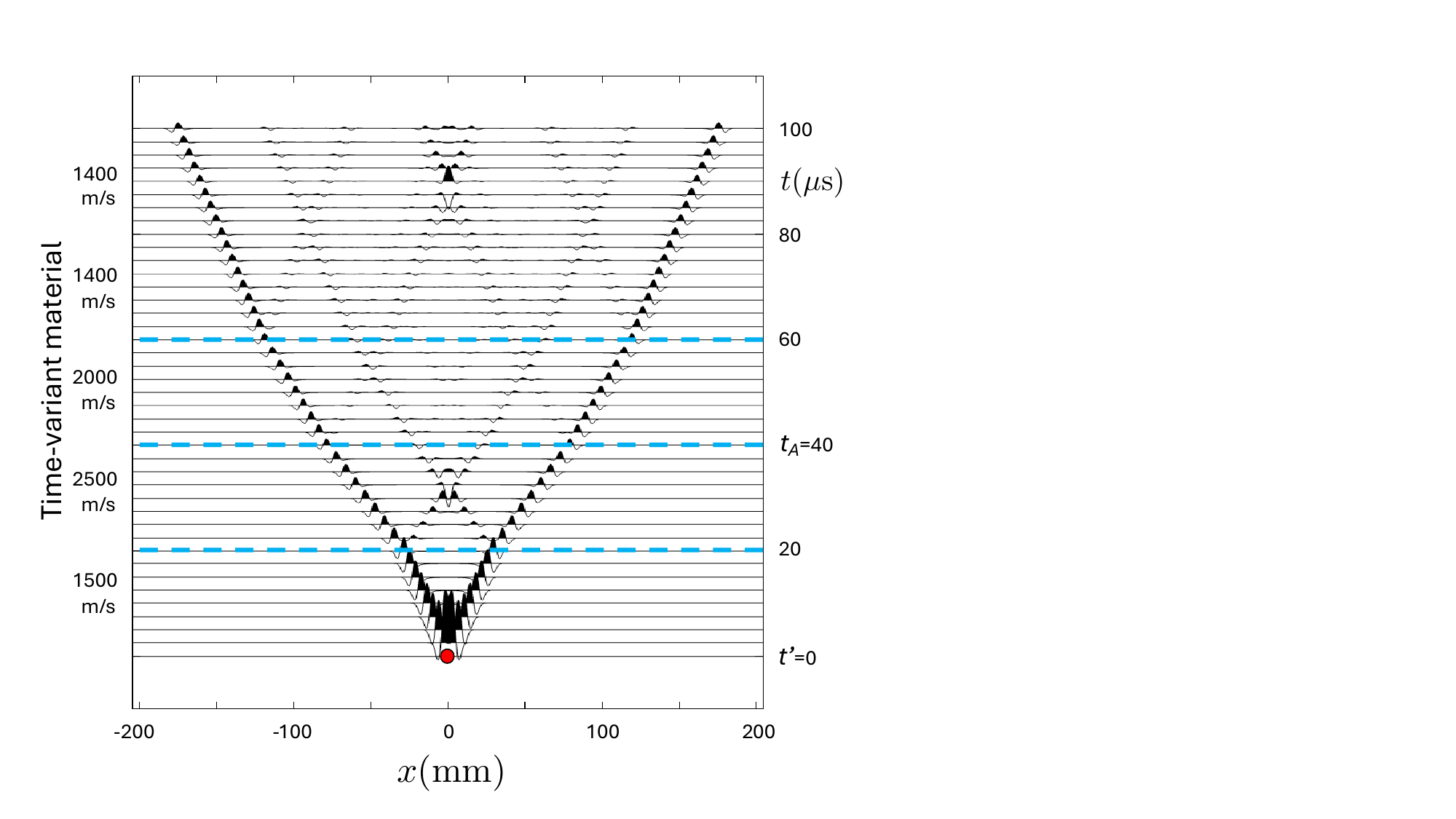}}
\caption{Cross-section of the Green's function of Figure \ref{Figure3} for $z=0$ as a function of $(x,t)$. The red dot indicates the source at ${\bf x}={\bf 0}$ and $t'=0$.
The dashed blue lines indicate the time boundaries.}\label{Figure5}
\end{figure}

We conclude this section with a 2D numerical example of a Green's function in a homogeneous, time-variant material.
We consider a piecewise constant material consisting of five time-invariant slabs, with propagation velocities of 1500, 2500, 2000, 1400 and 1400 m s$^{-1}$.
The parameter $\beta$ is taken constant throughout. The duration of each time-invariant slab is 20 $\mu$s. 
We use equation (\ref{eqrecur}), starting with $t_0=0$ $\mu$s, stepping through the material with steps of 2.5 $\mu$s.
For ${\bf W}({\bf x},t_n,t_{n-1})$ we use the analytical expression given in Appendix \ref{AppA2}. 
After each time step, we extract  the upper-right element, multiplied by $-1$, i.e., 
$-W^{U,V}({\bf x},t_n,t_0)={\cal G}_t({\bf x},{\bf 0},t_n,t_0)$ (the acausal Green's function is zero since $t_n>t_0$ for all $n$).
The result (convolved with a spatial wavelet with a central wavenumber $k_0/2\pi=100$ m$^{-1}$)
is shown in Figure \ref{Figure3} as a function of ${\bf x}=(x,z)$ and $t$ (for later convenience we have replaced $t_0$ by $t'=0$). 
Snapshots for constant $t$ are shown in Figure \ref{Figure4} and  
a cross-section as a function of $(x,t)$ for $z=0$ in Figure~\ref{Figure5}. These figures clearly show that when a wave encounters a time boundary 
(indicated by the blue dashed lines in Figure \ref{Figure5}), it is partly transmitted and partly reflected. 

\subsection{Homogeneous Green's function representation for a homogeneous, time-variant material}\label{sec4H}

As a counterpart of equation (\ref{eq1616}), which is a representation of the homogeneous Green's function ${\cal G}_x^h({\bf x},{\bf x}_A,t,0)$ in an inhomogeneous, time-invariant material,
here we seek for a representation of the homogeneous Green's function ${\cal G}_t^h({\bf x},{\bf 0},t,t_A)$ in a homogeneous, time-variant material.
To this end, we first formulate a representation of the propagator matrix ${\bf W}({\bf x},t,t_A)$.
As a special case of equation (\ref{eqrecur}) we write
\begin{eqnarray}
{\bf W}({\bf x},t,t_A)={\bf W}({\bf x},t,t')*_x{\bf W}({\bf x},t',t_A).\label{eqbasis}
\end{eqnarray}
In general, the order of $t$, $t'$ and $t_A$ is arbitrary, but in the following we assume that $t'$ is smaller than both $t_A$ and $t$.
Hence, on the right-hand side, the elements of matrix ${\bf W}({\bf x},t',t_A)$ (with $t'<t_A$) 
 consist only of (derivatives of) the acausal part of the homogeneous Green's functions.
On the other hand, the elements of matrix ${\bf W}({\bf x},t,t')$ (with $t'<t$) 
consist only of (derivatives of) the causal part of the homogeneous Green's functions.
Using the partitioning defined in equation (\ref{eq34}), we thus obtain for the upper-right element of ${\bf W}({\bf x},t,t_A)$ 
\begin{eqnarray}
W^{U,V}({\bf x},t,t_A)&=&W^{U,U}({\bf x},t,t')*_xW^{U,V}({\bf x},t',t_A)\nonumber\\
&+&W^{U,V}({\bf x},t,t')*_xW^{V,V}({\bf x},t',t_A),\label{eqWUV}
\end{eqnarray}
with $t'<t_A$ and $t'<t$, or, substituting equations (\ref{eq35aa}) -- (\ref{eq40}), 
\begin{eqnarray}
{\cal G}_t^h({\bf x},{\bf 0},t,t_A)&=&\beta(t')\Bigl[\{\partial_{t'}  {\cal G}_t({\bf x},{\bf 0},t,t')\}*_x{\cal G}_t^a({\bf x},{\bf 0},t',t_A)
\nonumber\\&&
-{\cal G}_t({\bf x},{\bf 0},t,t')*_x\partial_{t'}  {\cal G}_t^a({\bf x},{\bf 0},t',t_A)\Bigr].\label{eq58kk}
\end{eqnarray}
This is the homogeneous Green's function representation for a homogeneous, time-variant material. 
It is the counterpart of equation (\ref{eq1616}), with coordinates ${\bf x}$, ${\bf x}'$, ${\bf x}_A$ and $t$ replaced by $t$, $t'$, $t_A$ and ${\bf x}$.
Furthermore,  ${\bf \nabla}'$ is replaced by  $\partial_{t'}$ and $1/\beta({\bf x}')$ by $\beta(t')$. 
Whereas the temporal convolution in equation (\ref{eq1616}) is a 1D integral, 
and the integral over ${\bf x}'$ is  a 2D surface integral (for the 3D situation) or a 1D line integral (for the 2D situation), 
the spatial convolution in equation (\ref{eq58kk}) is a  3D or 2D integral, respectively, and there is no integral over $t'$ ($t'$ has only a single value in equation (\ref{eq58kk})).
Hence, the total number of dimensions of the integrals is the same in both representations.
Last, but not least, the time-reversed Green's function ${\cal G}_x({\bf x}',{\bf x}_A,-t,0)$ and its derivative in equation (\ref{eq1616}) are replaced
by the acausal Green's function ${\cal G}_t^a({\bf x},{\bf 0},t',t_A)$ and its derivative in equation (\ref{eq58kk}).

Whereas time-reversal can be applied to recordings of physical measurements, acausal responses cannot be obtained from physical measurements.
Therefore we need to modify equation (\ref{eq58kk}), or more generally, equation (\ref{eqbasis}), before we can use it in practical situations.
In particular, we need to use symmetry properties of the Green's functions
to transform equation (\ref{eqbasis}) into a form that complies with physics. In  section \ref{sec4D} we transform the  acausal matrix ${\bf W}({\bf x},t',t_A)$, 
propagating from $t_A$ to $t'$ (with $t'<t_A$), into a  causal
matrix, propagating from $-t_A$ to $-t'$ (with $-t'>-t_A$) in a time-reversed material. In this case, equation (\ref{eqbasis})  will form the theoretical basis for 
time-reversed acoustics.  In section \ref{sec4E} we transform the acausal matrix ${\bf W}({\bf x},t',t_A)$ into a causal matrix, propagating from $t'$ to $t_A$ (with $t_A>t'$).
In that case, equation (\ref{eqbasis})  will form the theoretical basis for Green's function retrieval by spatial correlation.

\subsection{Time-reversed acoustics in a homogeneous, time-variant material}\label{sec4D}

\begin{figure}[t]
\vspace{0cm}
\centerline{\hspace{14cm}\epsfxsize=24 cm \epsfbox{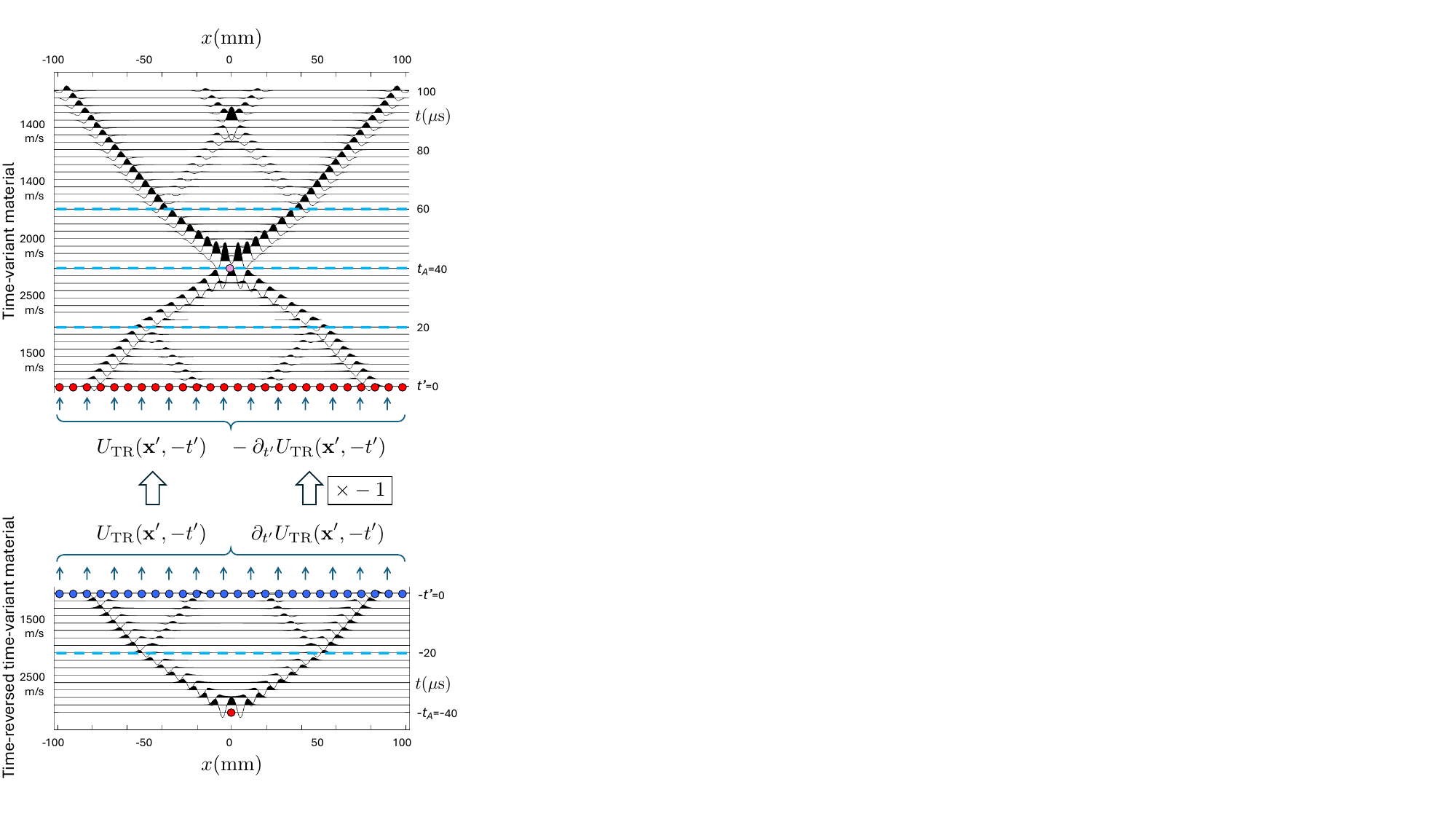}}
\caption{Illustration of the principle of time-reversed acoustics in a homogeneous, time-variant material. 
The dashed blue lines indicate the time boundaries.
Lower frame: the response $U_{\rm TR}({\bf x}',-t')$ (and its time-derivative $\partial_{t'}U_{\rm TR}({\bf x}',-t')$) 
to a source  at $-t_A=-40\,\mu$s and ${\bf x}={\bf 0}$ (indicated by the red dot) in the time-reversed version of the actual material, 
is recorded by receivers at $-t'=0\,\mu$s (only the response $U_{\rm TR}$ is shown).
Upper frame: after changing the sign of the time-derivative $\partial_{t'}U_{\rm TR}({\bf x}',-t')$, the recorded fields are fed to sources 
at all ${\bf x}'$ at $t'=0\,\mu$s.  The wave field emitted by these sources
 into the actual time-variant material focuses at $t_A=40\,\mu$s and ${\bf x}={\bf 0}$. After having focused, the wave field continues its propagation.
 The focused field at  $t_A=40\,\mu$s and ${\bf x}={\bf 0}$ is  a virtual source (indicated by the pink dot) for the field for $t>t_A$.
In section \ref{sec4E} we show that the upper frame can alternatively be obtained by the principle of Green's function retrieval in a homogeneous, time-variant material.}\label{Figure6}
\end{figure}

\begin{figure}[t]
\centerline{\hspace{4cm}\epsfxsize=13 cm \epsfbox{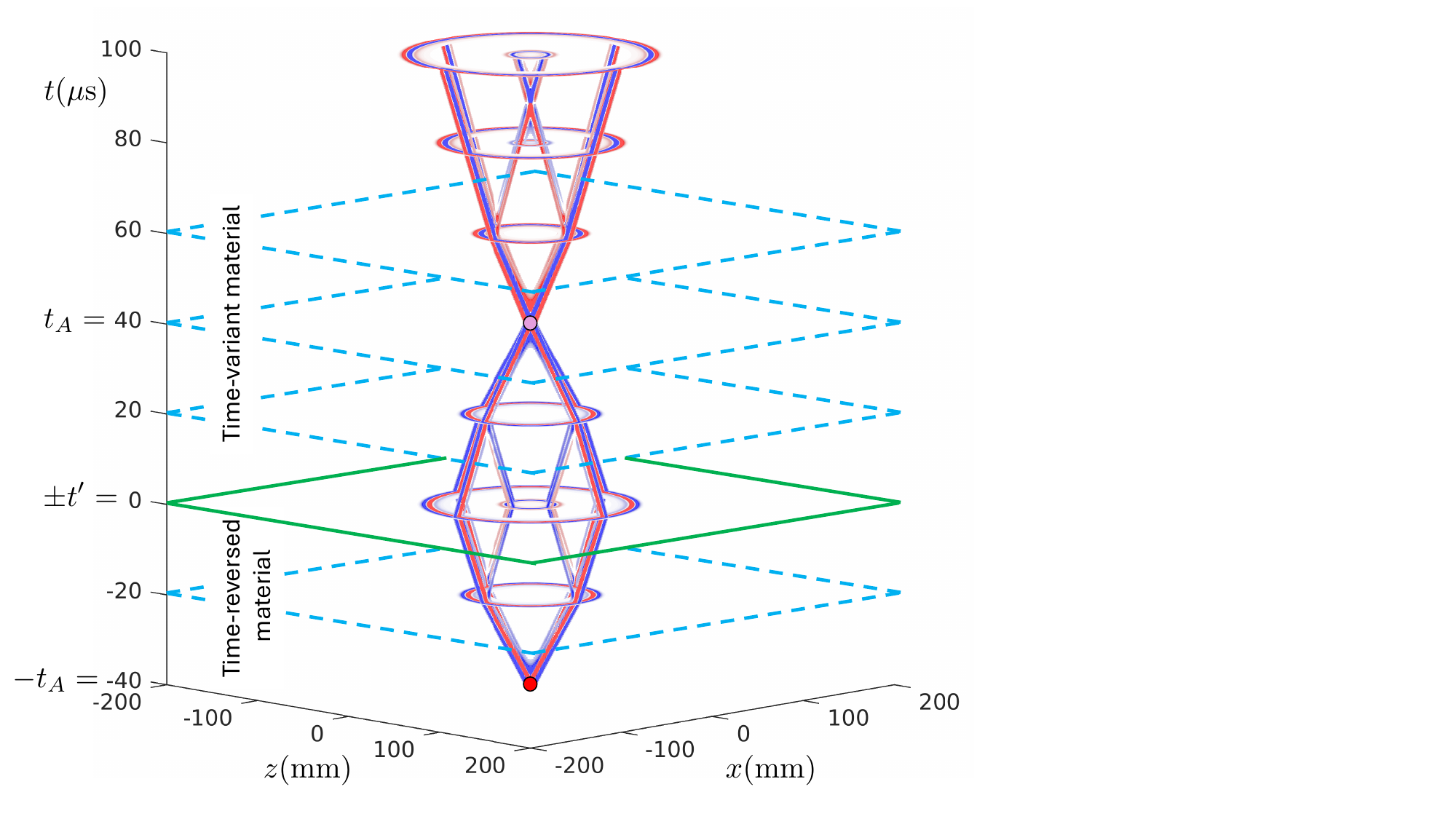}}
\caption{Wave field of Figure \ref{Figure6}, shown here as a function of ${\bf x}=(x,z)$ and $t$. 
The dashed blue planes indicate the time boundaries. The green plane indicates the interface between the actual material and its time-reversed version.
The pink dot at  $t_A=40\,\mu$s and ${\bf x}={\bf 0}$ indicates the virtual source.
Movie available at https://www.keeswapenaar.nl/TimeMaterial/TimeMirror.mp4}\label{Figure7}
\end{figure}

We start by introducing the principle of idealized time-reversed acoustics in a homogeneous, time-variant material in an intuitive way. We seek for a counterpart of 
time-reversed acoustics in an inhomogeneous, time-invariant material, as discussed in section \ref{sec3C}. Consider again Figures \ref{Figure1}(b) and \ref{Figure1}(c),
which illustrate a wave field, emitted by sources at ${\bf x}'$ on a closed boundary, propagating
through the inhomogeneous material, and focusing at  ${\bf x}_A$ and $t=0$ (Figure \ref{Figure1}(b)).
The focus at ${\bf x}_A$ is a virtual source, radiating waves into the inhomogeneous material (Figure \ref{Figure1}(c)). The counterpart of these figures is
shown in the upper frame of Figure \ref{Figure6}. Here we have sources at a single time-instant $t'=0$ $\mu$s, 
emitting waves into a homogeneous, time-variant material (the same material as used in the numerical example in section \ref{sec4C}). 
The field focuses at $t_A=40$ $\mu$s and ${\bf x}={\bf 0}$. 
The focus at $t_A$ is a virtual source, radiating waves into the time-variant material beyond $t_A$.
The question is, how do we obtain the wave field at $t'$, such that it focuses at $t_A$? Unlike in section \ref{sec3C}, where the 
field emitted into the inhomogeneous material was simply the time-reversal of the causal field, radiated by a source at ${\bf x}_A$ and $t=0$,
and observed at ${\bf x}'$  (Figure \ref{Figure1}(a)), in the present situation we cannot measure a field at $t'$, radiated by a source at $t_A>t'$.
Instead, we consider waves in a time-reversed version of the material between $t'$ and $t_A$. 
The lower frame of Figure \ref{Figure6} shows a wave field in the time-reversed material between $-t_A$ and $-t'$. A source at $-t_A$ and ${\bf x}={\bf 0}$
(indicated by the red dot) emits a wave field into
this time-reversed material. Note that the wave paths are opposite to those in the lower part of the upper frame. The wave field is observed
by receivers at $-t'>-t_A$ (indicated by the blue dots). 
These receivers measure the wave field $U_{\rm TR}({\bf x}',-t')$ and its time-derivative  $\partial_{t'}U_{\rm TR}({\bf x}',-t')$ 
(with subscripts ${\rm TR}$ denoting that these fields are measured in the time-reversed material; for simplicity Figure \ref{Figure6} only shows the response $U_{\rm TR}$). 
To let the waves change their propagation direction, we need to change the sign of $\partial_{t'}U_{\rm TR}({\bf x}',-t')$ \cite{Bacot2016NP, Fink2017EPJ}.
Hence, $U_{\rm TR}({\bf x}',-t')$ and  $-\partial_{t'}U_{\rm TR}({\bf x}',-t')$ form the field that needs to be fed to the sources
at all ${\bf x}'$ at $t'$ and emitted into the actual time-variant material, to focus at 
$t_A$ and ${\bf x}={\bf 0}$. Note that Figure \ref{Figure6} can be seen as a cross-section for $z=0$ of the wave field in Figure \ref{Figure7}.

Next, we formalize this principle of idealized time-reversed acoustics in a homogeneous, time-variant material, using the representation of equation (\ref{eqbasis}) as starting point.
The acausal Green's functions in matrix ${\bf W}({\bf x},t',t_A)$, with $t'<t_A$,  are not associated to a physical situation. 
We use symmetry properties of the Green's functions  to transform ${\bf W}({\bf x},t',t_A)$ in equation (\ref{eqbasis}) into a form that complies with physics. 

Consider the Green's function ${\cal G}_t({\bf x},{\bf 0},t,t_A)$, obeying wave equation (\ref{eq31agbl}), with the causality condition of equation (\ref{eq31con})
(for ${\bf x}_A={\bf 0}$).
We introduce ${\cal G}_{t,{\rm TR}}({\bf x},{\bf 0},t,-t_A)$ as the response to a source at $-t_A$ in the time-reversed version of the actual material.
Hence, it obeys the following equation
\begin{eqnarray}
\bigl(\partial_t\beta(-t)\partial_t - \beta(-t)c^2(-t){\bf \nabla}^2\bigr){\cal G}_{t,{\rm TR}}({\bf x},{\bf 0},t,-t_A)= \delta({\bf x})\delta(t+t_A),\label{eq31agbltr}
\end{eqnarray}
with causality condition
\begin{eqnarray}
{\cal G}_{t,{\rm TR}}({\bf x},{\bf 0},t,-t_A)=0 \quad \mbox{for}\quad t<-t_A.\label{eq31contr}
\end{eqnarray}
Note that the subscript TR denotes that the material is time-reversed (hence, TR does not refer to time-reversal of the wave field).
Replacing $t$ by $-t$ in equation (\ref{eq31agbltr}) gives
\begin{eqnarray}
\bigl(\partial_t\beta(t)\partial_t - \beta(t)c^2(t){\bf \nabla}^2\bigr){\cal G}_{t,{\rm TR}}({\bf x},{\bf 0},-t,-t_A)= \delta({\bf x})\delta(-t+t_A),\label{eq31agbltrm}
\end{eqnarray}
with causality condition
\begin{eqnarray}
{\cal G}_{t,{\rm TR}}({\bf x},{\bf 0},-t,-t_A)=0 \quad \mbox{for}\quad -t<-t_A.\label{eq31contrm}
\end{eqnarray}
Note that $\delta(-t+t_A)$ in equation (\ref{eq31agbltrm}) is equal to $\delta(t-t_A)$, and the condition $-t<-t_A$ in equation (\ref{eq31contrm}) is equivalent with
the condition $t>t_A$. Hence, ${\cal G}_{t,{\rm TR}}({\bf x},{\bf 0},-t,-t_A)$ obeys the same wave equation (equation (\ref{eq31agbl}), with ${\bf x}_A={\bf 0}$) 
and the same causality condition (equation (\ref{eq31conac})) as ${\cal G}_t^a({\bf x},{\bf 0},t,t_A)$.
Hence,
\begin{eqnarray}
{\cal G}_t^a({\bf x},{\bf 0},t,t_A)={\cal G}_{t,{\rm TR}}({\bf x},{\bf 0},-t,-t_A).\label{eq5656}
\end{eqnarray}
Similarly,
\begin{eqnarray}
{\cal G}_t({\bf x},{\bf 0},t,t_A)&=&{\cal G}_{t,{\rm TR}}^a({\bf x},{\bf 0},-t,-t_A),\\
{\cal G}_t^h({\bf x},{\bf 0},t,t_A)&=&-{\cal G}_{t,{\rm TR}}^h({\bf x},{\bf 0},-t,-t_A),\label{eqGhTR}
\end{eqnarray}
where ${\cal G}_{t,{\rm TR}}^a$ and ${\cal G}_{t,{\rm TR}}^h$ are the acausal and homogeneous Green's functions, respectively, in the time-reversed material.
Using equation (\ref{eqGhTR}) in equations (\ref{eq35aa})--(\ref{eq41}) we find
\begin{eqnarray}
W^{U,V}({\bf x},t,t_A)&=&- W_{\rm TR}^{U,V}({\bf x},-t,-t_A),\\
W^{V,V}({\bf x},t,t_A)&=&W_{\rm TR}^{V,V}({\bf x},-t,-t_A),\\
W^{U,U}({\bf x},t,t_A)&=&W_{\rm TR}^{U,U}({\bf x},-t,-t_A),\\
W^{V,U}({\bf x},t,t_A)&=&- W_{\rm TR}^{V,U}({\bf x},-t,-t_A).
\end{eqnarray}
Hence,
\begin{eqnarray}
{\bf W}({\bf x},t,t_A)={\bf J}{\bf W}_{\rm TR}({\bf x},-t,-t_A){\bf J}^{-1},\label{eq47}
\end{eqnarray}
where ${\bf W}_{\rm TR}({\bf x},-t,-t_A)$ is the propagator matrix for the time-reversed material, and
\begin{eqnarray}
{{\bf J}}=\begin{pmatrix} 1 & 0 \\ 0 & -1 \end{pmatrix}.\label{eq48}
\end{eqnarray}
Substituting equation (\ref{eq47}) into equation (\ref{eqbasis}) and post-multiplying both sides by ${\bf J}$ yields
\begin{eqnarray}
{\bf W}({\bf x},t,t_A){\bf J}={\bf W}({\bf x},t,t')*_x{\bf J}{\bf W}_{\rm TR}({\bf x},-t',-t_A).\label{eqbasistr}
\end{eqnarray}
In equation (\ref{eqbasistr}), ${\bf W}_{\rm TR}({\bf x},-t',-t_A)$ propagates from $-t_A$ to $-t'$ through the time-reversed version of the actual material between
$t'$ and $t_A$.  Hence, the elements of both propagator matrices on the right-hand side of equation (\ref{eqbasistr})
consist only of (derivatives of) the causal part of the homogeneous Green's functions.
Let us compare equation (\ref{eqbasistr}) with its counterpart of equation (\ref{eq1515}). To this end, 
for the upper-right element of ${\bf W}({\bf x},t,t_A){\bf J}$ we write
\begin{eqnarray}
-W^{U,V}({\bf x},t,t_A)&=&W^{U,U}({\bf x},t,t')*_xW_{\rm TR}^{U,V}({\bf x},-t',-t_A)\nonumber\\
&&-W^{U,V}({\bf x},t,t')*_xW_{\rm TR}^{V,V}({\bf x},-t',-t_A).
\end{eqnarray}
Substituting equations (\ref{eq35aa}) -- (\ref{eq40}), using  $\beta_{\rm TR}(-t')=\beta(t')$, yields
\begin{eqnarray}
{\cal G}_t^h({\bf x},{\bf 0},t,t_A) 
&=&\beta(t')\Bigl[\{\partial_{t'}  {\cal G}_t({\bf x},{\bf 0},t,t')\}*_x{\cal G}_{t, {\rm TR}}({\bf x},{\bf 0},-t',-t_A)\nonumber\\
&&-{\cal G}_t({\bf x},{\bf 0},t,t')*_x\partial_{t'}  {\cal G}_{t, {\rm TR}}({\bf x},{\bf 0},-t',-t_A)\Bigr].\label{eq58gg}
\end{eqnarray}
This expression is the counterpart of equation (\ref{eq1515}). Unlike in equation (\ref{eq1515}), where we captured the two terms under the integral in equation (\ref{eq1616})
by a single term, no such simplification is possible for equation (\ref{eq58gg}).

We continue with the matrix representation of equation (\ref{eqbasistr}) and use this as the basis for  time-reversed acoustics in a homogeneous, time-variant material.
To this end, we define a source vector ${\bf d}({\bf x},t)$ at $t=-t_A$ in the time-reversed material, according to ${\bf d}({\bf x},t)={\bf d}_0({\bf x})\delta(t+t_A)$. Then 
the causal response to this source in the time-reversed material is given by
\begin{eqnarray}
{\bf q}_{\rm TR}({\bf x},t)&=&{\bf W}_{\rm TR}({\bf x},t,-t_A)*_x{\bf d}_0({\bf x}).\label{eqA22ag}
\end{eqnarray}
Convolving both sides of equation (\ref{eqbasistr}) with the source distribution ${\bf d}_0({\bf x})$, using equation (\ref{eqA22ag}) for $t=-t'$, yields
\begin{eqnarray}
{\bf W}({\bf x},t,t_A)*_x{\bf J}{\bf d}_0({\bf x})={\bf W}({\bf x},t,t')*_x{\bf J}{\bf q}_{\rm TR}({\bf x},-t').\label{eqbasistrsrc}
\end{eqnarray}
We discuss this expression from right to left. For this discussion, recall that $t_A$ and $t'$ are fixed, whereas $t$ is a variable; 
moreover, the order of $t_A$ and $t$ is arbitrary and $t'$ is smaller than both $t_A$ and $t$. The term 
\begin{eqnarray}
{\bf q}_{\rm TR}({\bf x},-t')=\begin{pmatrix}U_{\rm TR}({\bf x},-t')\\V_{\rm d, TR}({\bf x},-t')\end{pmatrix}\label{eq52a}
\end{eqnarray}
is, according to equation (\ref{eqA22ag}), the causal response to the source at $-t_A$ in the time-reversed material, observed at $-t'$, 
see the lower frame of Figure \ref{Figure6} (recall that for simplicity Figure \ref{Figure6} only shows the component $U_{\rm TR}$).
According to equation (\ref{eq49b}) (with $a=0$) we have $V_{\rm d,TR}({\bf x},-t')=\beta(t')\partial_{t'}U_{\rm TR}({\bf x},-t')$.
The matrix ${\bf J}$ in the right-hand side of equation (\ref{eqbasistrsrc})  changes the sign of the lower component of ${\bf q}_{\rm TR}({\bf x},-t')$, hence
\begin{eqnarray}
{\bf J}{\bf q}_{\rm TR}({\bf x},-t')=\begin{pmatrix}U_{\rm TR}({\bf x},-t')\\
-\beta(t')\partial_{t'}U_{\rm TR}({\bf x},-t')
\end{pmatrix}.\label{eq53a}
\end{eqnarray}
This sign change is indicated between the two frames in Figure \ref{Figure6}.
Next, matrix ${\bf W}({\bf x},t,t')$ in equation (\ref{eqbasistrsrc})  propagates this field from $t'$ 
through the actual time-variant material to $t$, see the upper frame of Figure \ref{Figure6}.
For $t<t_A$, the left-hand side of equation (\ref{eqbasistrsrc}) is interpreted as an acausal field observed at $t$, propagating to a sink at $t_A$. For $t>t_A$, 
it is interpreted as a causal field observed at $t$, radiated by a  source at $t_A$ of strength  ${\bf J}{\bf d}_0({\bf x})$.
Since there is no real sink nor a real source at $t_A$, the sink and the source at $t_A$ in the interpretation of the left-hand side of equation (\ref{eqbasistrsrc}) are both virtual.
In the numerical example in Figures \ref{Figure6} and \ref{Figure7}, the virtual source at $t_A=40 \mu$s coincides with a time boundary, but note that the theory described here
holds for any choice of $t_A$.

Summarizing, equation (\ref{eqbasistrsrc}) formalizes the principle of idealized time-reversed acoustics  in a homogeneous, time-variant material.
To compare this expression with equation (\ref{eq1717}), we choose for convenience
\begin{eqnarray}
{\bf d}_0({\bf x})=\begin{pmatrix} 0\\b_{{\rm d},0}({\bf x})\end{pmatrix}.\label{eq67}
\end{eqnarray}
For the upper element of equation (\ref{eqbasistrsrc}) we thus obtain (using equation (\ref{eq53a}))
\begin{eqnarray}
-W^{U,V}({\bf x},t,t_A)*_xb_{{\rm d},0}({\bf x})&=&W^{U,U}({\bf x},t,t')*_xU_{\rm TR}({\bf x},-t')\nonumber\\
&+&W^{U,V}({\bf x},t,t')*_x\{-\beta(t')\partial_{t'}U_{\rm TR}({\bf x},-t')\},
\end{eqnarray}
or, using equations (\ref{eq35aa}) and (\ref{eq40}) and (a)causality conditions (\ref{eq31con}) and (\ref{eq31conac}), 
\begin{eqnarray}
\{{\cal G}_t({\bf x},{\bf 0},t,t_A) - {\cal G}_t^a({\bf x},{\bf 0},t,t_A)\}*_xb_{{\rm d},0}({\bf x})&=&
-\beta(t')\Bigl[\{\partial_{t'}  {\cal G}_t({\bf x},{\bf 0},t,t')\}*_x{U_{\rm TR}({\bf x},-t')}\nonumber\\
&&+{\cal G}_t({\bf x},{\bf 0},t,t')*_x{\{-\partial_{t'}U_{\rm TR}({\bf x},-t')\}}\Bigr].\label{eq58gghh}
\end{eqnarray}
Using the integral notation for the spatial convolutions at the right-hand side and  the space-shift invariance of the Green's function (equation (\ref{eq31shf})), this becomes
\begin{eqnarray}
&&\hspace{-1cm}\int_{\mathbb{R}^m}\{{\cal G}_t({\bf x},{\bf x}',t,t_A) - {\cal G}_t^a({\bf x},{\bf x}',t,t_A)\}b_{{\rm d},0}({\bf x}'){\rm d}{\bf x}'=\nonumber\\
&&\hspace{.0cm}-\beta(t')\int_{\mathbb{R}^m}\Bigl[
\underbrace{\{\partial_{t'}  {\cal G}_t({\bf x},{\bf x}',t,t')\}}_{\mbox{\footnotesize``propagator''}}
\underbrace{U_{\rm TR}({\bf x}',-t')}_{\mbox{\footnotesize``source field''}}\nonumber\\
&&\hspace{2cm}+
\underbrace{{\cal G}_t({\bf x},{\bf x}',t,t')}_{\mbox{\footnotesize``propagator''}}
\underbrace{\{-\partial_{t'}U_{\rm TR}({\bf x}',-t')\}}_{\mbox{\footnotesize``source field''}}\Bigr]{\rm d}{\bf x}'.\label{eq58gghhb}
\end{eqnarray}
This expression is the counterpart of equation (\ref{eq1717}).
The time-reversed field $P({\bf x}',-t)$ in equation (\ref{eq1717}) is replaced by the fields $U_{\rm TR}({\bf x}',-t')$ and $-\partial_{t'}U_{\rm TR}({\bf x}',-t')$ 
in a time-reversed version of the actual material. 
The right-hand side quantifies the propagation of $U_{\rm TR}({\bf x}',-t')$ and $-\partial_{t'}U_{\rm TR}({\bf x}',-t')$ from sources at all ${\bf x}'$ at 
a single time instant $t'$ to any $t$ larger than $t'$.
Unlike in  equation (\ref{eq1717}), the two terms on the right-hand side of equation (\ref{eq58gghh}) cannot be recast into a single term.
The second term on the left-hand side, $\int_{\mathbb{R}^m}{\cal G}_t^a({\bf x},{\bf x}',t,t_A)b_{{\rm d},0}({\bf x}'){\rm d}{\bf x}'$, is the field that focuses at $t_A$,
and the first term on the left-hand side, $\int_{\mathbb{R}^m}{\cal G}_t({\bf x},{\bf x}',t,t_A)b_{{\rm d},0}({\bf x}'){\rm d}{\bf x}'$, is the field emitted by
the virtual source distribution $b_{{\rm d},0}({\bf x})$ at $t_A$, see the upper frame of Figure \ref{Figure6}. 

What we have discussed in this section is the ideal version of time-reversed acoustics in a homogeneous, time-variant material.
It requires a wave field in a time-reversed version of the actual material, measurements of this field and its time derivative at all ${\bf x}'$ at a single time-instant $-t'$, and emission
of this field and its time derivative (the latter with reversed sign) into the actual material by sources at all  ${\bf x}'$ at time-instant $t'$. 
Experiments described by references \cite{Bacot2016NP, Fink2017EPJ, Hidalgo2023PRL, Peng2020JPC} make use of several shortcuts.
These experiments are carried out in a material that is time-invariant, except for an impulsive temporal disruption at $t'=0$. 
Hence, the time-reversed material is the same as the actual material. Furthermore,
instead of measuring the field and its time derivative and re-emitting this (with a sign change) into the material, the impulsive temporal disruption of the material 
at $t'=0$ acts as a time boundary, which partly reflects and partly transmits the field into the material. 
The reflected field approximates the re-emitted field and focuses at the position of the original source(s),  whereas the transmitted field, which continues propagating forward, is ignored.
Despite the approximations, these experiments are fascinating and triggered the analysis described in this paper.

\subsection{Green's function retrieval in a homogeneous, time-variant material}\label{sec4E}

\begin{figure}[t]
\centerline{\hspace{0cm}\epsfxsize=16 cm \epsfbox{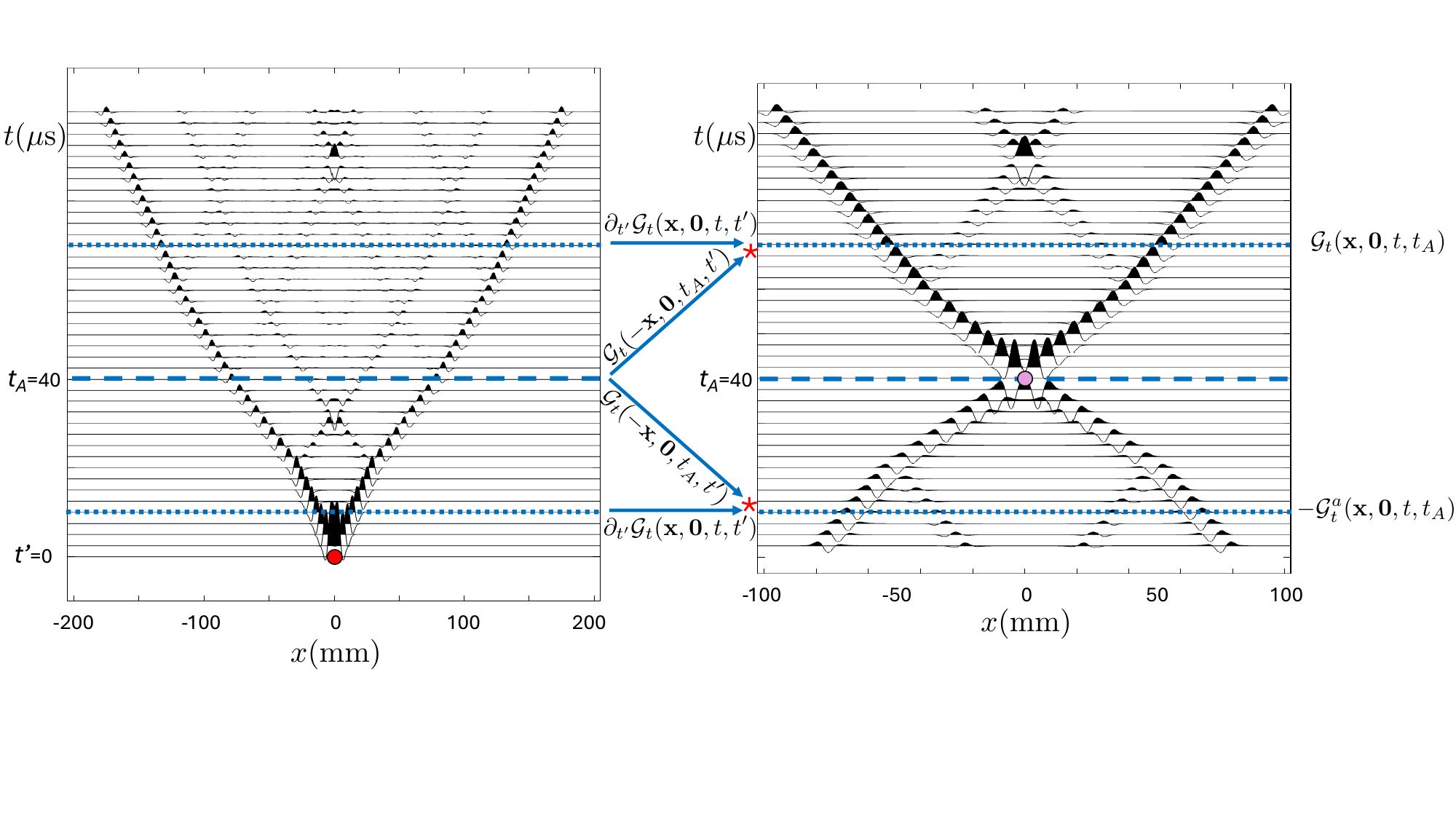}}
\vspace{-2cm}
\caption{Illustration of the principle of Green's function retrieval in a homogeneous, time-variant material.
The left frame shows the Green's function ${\cal G}_t({\bf x},{\bf 0},t,t'=0)$  (the same as in Figure \ref{Figure5}), with its source indicated by the red dot. 
The spatial crosscorrelation of the response  at $t_A=40\,\mu$s with that at any other $t$
(explained in more detail in the text), yields the response to a virtual source, indicated
by the pink dot in the right frame (note the different scales of the space axes). The response in the right frame is identical to that in the upper frame of Figure \ref{Figure6}, 
which was obtained by the principle of time-reversed acoustics, discussed in section \ref{sec4D}.
}\label{Figure8}
\end{figure}

We start by introducing the principle of Green's function retrieval in a homogeneous, time-variant material in an intuitive way.
We seek for a counterpart of Green's function retrieval in an  inhomogeneous, time-invariant material, as discussed in section \ref{sec3D}.
Consider again Figure \ref{Figure2}(a), which illustrates Green's functions in an inhomogeneous material, with impulsive sources at $t=0$ and ${\bf x}'$  on a closed boundary, 
and receivers at ${\bf x}_A$ and ${\bf x}$ inside the closed boundary. The temporal crosscorrelation of the Green's functions, observed at ${\bf x}_A$ and ${\bf x}$, 
integrated over all sources at ${\bf x}'$, yields the Green's function (minus its time-reversed version) between ${\bf x}_A$ and ${\bf x}$.
Hence, the receiver at ${\bf x}_A$ has turned into a virtual source, and the response to this source is observed at ${\bf x}$,
see Figure \ref{Figure2}(b). The counterpart of Figure  \ref{Figure2}(a) is shown in the left frame of Figure \ref{Figure8}.
It shows the Green's function ${\cal G}_t({\bf x},{\bf 0},t,t'=0)$ (convolved with a spatial wavelet) in a time-variant material, with its source at ${\bf x}={\bf 0}$ and $t'=0$
(indicated by the red dot), and receivers at $t_A=40\,\mu$s 
(indicated by the dashed blue line) and variable $t$ (two of them indicated by the dotted blue lines).
We expect that the spatial crosscorrelation of the Green's function observed at $t_A=40\,\mu$s (the dashed blue line), 
with that at any other  $t$ (the dotted blue lines), yields the Green's function (minus its acausal version)
between $t_A$ and $t$, as shown in the right frame of Figure \ref{Figure8}. This frame can be interpreted as the counterpart of Figure \ref{Figure2}(b).
It shows the response to a virtual source at ${\bf x}={\bf 0}$ and $t_A=40\,\mu$s, indicated by the pink dot, observed by receivers at variable ${\bf x}$ and $t$
(the dotted blue lines). 
In the following we show mathematically that this Green's function (minus its acausal version) indeed follows from spatial crosscorrelations of Green's functions,
albeit in a somewhat more complex way than described above.

Our starting point is again the representation of equation (\ref{eqbasis}). The acausal Green's functions in matrix ${\bf W}({\bf x},t',t_A)$, with $t'<t_A$,
are not associated to a physical situation.
We use symmetry properties of the Green's functions  to transform this matrix into a form that complies with physics. 
Using reciprocity relation (\ref{eq1022sqshblkkk}) in equations (\ref{eq35aa})--(\ref{eq41}) we find
\begin{eqnarray}
W^{U,V}({\bf x},t,t_A)&=&- W^{U,V}({\bf x},t_A,t),\\
W^{V,V}({\bf x},t,t_A)&=&W^{U,U}({\bf x},t_A,t),\\
W^{U,U}({\bf x},t,t_A)&=&W^{V,V}({\bf x},t_A,t),\\
W^{V,U}({\bf x},t,t_A)&=&- W^{V,U}({\bf x},t_A,t).
\end{eqnarray}
Hence
\begin{eqnarray}
{\bf W}({\bf x},t,t_A)={\bf N}{\bf W}^t({\bf x},t_A,t){\bf N}^{-1},\label{eq253c}
\end{eqnarray}
where superscript $t$ denotes transposition and
\begin{eqnarray}
{{\bf N}}=\begin{pmatrix} 0 & 1 \\ -1 & 0 \end{pmatrix}.\label{eq48N}
\end{eqnarray}
Substituting equation (\ref{eq253c}) into equation (\ref{eqbasis}) and post-multiplying both sides by ${\bf N}$, we obtain 
\begin{eqnarray}
{\bf W}({\bf x},t,t_A){\bf N}={\bf W}({\bf x},t,t')*_x{\bf N}{\bf W}^t({\bf x},t_A,t').\label{eqbasisbb}
\end{eqnarray}
The right-hand side represents a spatial convolution. From equations (\ref{eq68blue})--(\ref{eq33}) it follows that the propagator matrices are symmetric in ${\bf x}$.
Hence,  we may replace ${\bf x}$ in the rightmost matrix in equation (\ref{eqbasisbb}) by $-{\bf x}$, so that the right-hand side becomes a spatial crosscorrelation.
Then, for the upper-left element we obtain
\begin{eqnarray}
-W^{U,V}({\bf x},t,t_A)&=&W^{U,U}({\bf x},t,t')*_xW^{U,V}(-{\bf x},t_A,t')\nonumber\\
&-&W^{U,V}({\bf x},t,t')*_xW^{U,U}(-{\bf x},t_A,t').
\end{eqnarray}
We assume again that the order of $t_A$ and $t$ is arbitrary and that $t'$ is smaller than both $t_A$ and $t$.
Using equations (\ref{eq35aa}) and (\ref{eq40}) and (a)causality conditions (\ref{eq31con}) and (\ref{eq31conac}), we thus obtain
\begin{eqnarray}
{\cal G}_t({\bf x},{\bf 0},t,t_A) - {\cal G}_t^a({\bf x},{\bf 0},t,t_A)&=&
\beta(t')\Bigl[\{\partial_{t'}  {\cal G}_t({\bf x},{\bf 0},t,t')\}*_x{\cal G}_t(-{\bf x},{\bf 0},t_A,t')\nonumber\\
&&-{\cal G}_t({\bf x},{\bf 0},t,t')*_x\partial_{t'}  {\cal G}_t(-{\bf x},{\bf 0},t_A,t')\Bigr].\label{eq58}
\end{eqnarray}
This expression is the counterpart of equation (\ref{eq1919}).
The right-hand side quantifies the spatial crosscorrelation of specific combinations of Green's functions and their derivatives, 
observed  at all ${\bf x}$ at $t_A$ and $t$, in response to sources at a single time instant $t'$.
Unlike in equation (\ref{eq22}), the two terms on the right-hand side of equation (\ref{eq58}) cannot be recast into a single term with a single source type,
meaning that we require responses to two types of sources at $t'$.
The left-hand side is the retrieved Green's function between $t_A$ and $t$ (minus its acausal version), see the right frame of Figure \ref{Figure8},
which is identical to the upper frame of Figure \ref{Figure6} in section \ref{sec4D}.
Whereas in section \ref{sec4D} this was obtained by emitting the response of a time-reversed material into the actual time-variant material (equation(\ref{eq58gghh})
and Figure \ref{Figure6}),
here it is explained as the result of crosscorrelating responses at two time instances in one-and-the-same material (equation (\ref{eq58}) and Figure \ref{Figure8}). 
Both methods were derived from equation (\ref{eqbasis}) and a specific reorganization of the matrix ${\bf W}({\bf x},t',t_A)$.
This confirms the link between  time-reversal acoustics and Green's function retrieval by spatial crosscorrelation in a homogeneous, time-variant material. 

The modification of equation (\ref{eq58}) for Green's function retrieval by spatial crosscorrelation of ambient noise observations at $t_A$ and $t$
is more complicated than in section \ref{sec3D}, because the two terms on the right-hand side of equation (\ref{eq58}) cannot be recast into a single term with a single source type.
In a companion paper \cite{Aichele2025arXiv} we discuss a compromise for the special case of a time-invariant material,
and discuss its relation with the SPAC (spatial autocorrelation) method \cite{Aki57BERI, Cox73JASA, Asten2006GEO}.
Extending this approach to time-variant materials leads to artifacts in the retrieved Green's function that increase with increasing complexity of the time-variant material. 
A further discussion of Green's function retrieval by spatial crosscorrelation of ambient noise is beyond the scope of this paper.

\section{Conclusions}\label{sec5}

We started this paper with a review of the principles of classical time-reversed acoustics and Green's function retrieval for inhomogeneous, time-invariant materials.
Along the same lines, we discussed the counterparts of these principles for homogeneous, time-variant materials, and compared those with the classical principles.
One observation from this comparison is that the space and time coordinates are interchanged between the methods for the two classes of material. 
As a consequence, temporal convolutions and correlations in the classical approaches for inhomogeneous, time-invariant materials, 
are replaced by spatial convolutions and correlations in their counterparts for homogeneous, time-variant materials. However, the interchangement of space and time 
does not apply to causality conditions, which are always related to the time coordinate. 
This complicates the translation of methods from one class of material to the other and vice versa.
Another complicating factor is that the number of time dimensions (one) is not the same as the number of space dimensions (two or three). 

For time-reversed acoustics and for Green's function retrieval, we separately review the relations between applications in the two classes of material.

\begin{itemize}
\item The principle of classical time-reversed acoustics in an inhomogeneous, time-invariant material involves in essence: 
\begin{enumerate}
\item Emission of a wave field by a source at ${\bf x}_A$ into the inhomogeneous material.
\item Recording of this wave field by receivers on a closed boundary.
\item Emission of the time-reversal of the recorded field into the inhomogeneous material by sources at the positions of the receivers.
\item The emitted wave field focuses at ${\bf x}_A$ and the focused field is a virtual source, emitting waves into the inhomogeneous material.
\end{enumerate}

The counterpart of the principle of classical time-reversed acoustics for a homogeneous, time-variant material involves:
\begin{enumerate}
\item Emission of a wave field by a source at $-t_A$ into a time-reversed version of the actual time-variant material. 
\item Recording of this wave field and its time-derivative by receivers at a single time instant.
\item Emission of the recorded field and its time-derivative (the latter with reversed sign) into the actual time-variant material by sources at the positions of the receivers.
\item The emitted wave field focuses at $t_A$ and the focused field is a virtual source, emitting waves into the time-variant material.
\end{enumerate}
This summarizes ideal time-reversed acoustics in a homogeneous, time-variant material.
In actual experiments  \cite{Bacot2016NP, Fink2017EPJ, Hidalgo2023PRL, Peng2020JPC}, the material is time-invariant except for an impulsive
temporal disruption. This disruption acts as a time boundary, which partly reflects the field into the material and as such avoids recording and re-emission of the wave field.

\item The principle of classical Green's function retrieval in an inhomogeneous, time-invariant material involves in essence:
\begin{enumerate}
\item Recording of passive wave fields at ${\bf x}_A$ and ${\bf x}$ in response to sources at ${\bf x}'$ on a closed boundary.
\item Temporal crosscorrelating the recorded fields at ${\bf x}_A$ and ${\bf x}$ per source, followed by a summation over all sources on the closed boundary.
\item The time derivative of the result is the Green's function (minus its time-reversed version) between a virtual source (sink) at ${\bf x}_A$ and a physical receiver at ${\bf x}$.
\item In the case that the sources are simultaneously acting uncorrelated noise sources, step 2 is replaced 
by a single crosscorrelation of the ambient noise recordings at ${\bf x}_A$ and ${\bf x}$.
\end{enumerate}

The counterpart of the principle of classical Green's function retrieval for a homogeneous, time-variant material involves:
\begin{enumerate}
\item Recording of passive wave fields at $t_A$ and $t$ in response to two types of sources at a single time instant $t'$.
\item Spatial crosscorrelating two specific combinations of the recorded fields at $t_A$ and $t$ per source type, followed by a summation of the two terms.
\item The result is the Green's function (minus its acausal version) between a virtual source (sink) at $t_A$ and a physical receiver at $t$.
\item A modification of Green's function retrieval by spatial crosscorrelation in the case of simultaneously acting uncorrelated noise sources has thus far only been derived for a
time-invariant material \cite{Aichele2025arXiv}.
\end{enumerate}
\end{itemize}

Finally, we review the relations between time-reversed acoustics and Green's function retrieval, first for an inhomogeneous, time-invariant material, and after that for
a homogeneous, time-variant material.
\begin{itemize}
\item For an inhomogeneous, time-invariant material, the principles of time-reversed acoustics and Green's function retrieval are variations
of the homogeneous Green's function representation of equation (\ref{eq1515}) for this class of material.
In particular, consider the temporal convolutional product 
\begin{eqnarray}
\{{\bf \nabla}'{\cal G}_x({\bf x},{\bf x}',t,0)\cdot{\bf n}\}*_t{\cal G}_x({\bf x}',{\bf x}_A,-t,0)\label{eqconprod1}
\end{eqnarray}
under the integral in equation (\ref{eq1515}).
\begin{enumerate}
\item In time-reversed acoustics, the second Green's function (after a temporal convolution with a source wavelet) stands for the time-reversed field $P({\bf x}',-t)$. This field is fed
to sources at ${\bf x}'$ on the closed boundary ${\cal S}$ and propagated by the first Green's function from ${\bf x}'$ to any  point ${\bf x}$ inside ${\cal S}$, 
as illustrated in Figure \ref{Figure1}b. The result, after integration over ${\cal S}$, is the homogeneous Green's function ${\cal G}_x^h({\bf x},{\bf x}_A,t,0)$
(temporally convolved with the time-reversal of the source wavelet). This is interpreted as a field focusing at ${\bf x}_A$ (Figure \ref{Figure1}b) and, after having focussed,
propagating away from ${\bf x}_A$ (Figure \ref{Figure1}c).
\item In Green's function retrieval, the second Green's function in equation (\ref{eqconprod1}) is, 
on basis of reciprocity relation (\ref{eq31shfrec}), replaced by ${\cal G}_x({\bf x}_A,{\bf x}',-t,0)$. Now the temporal convolutional product
can be interpreted as  a temporal crosscorrelation (due to the time-reversal of the second Green's function) of observations by receivers at ${\bf x}$ and ${\bf x}_A$,
in response to sources at ${\bf x}'$ on ${\cal S}$, as illustrated in Figure \ref{Figure2}a.
The result, after integration over ${\cal S}$, is the homogeneous Green's function ${\cal G}_x^h({\bf x},{\bf x}_A,t,0)$, which (for $t>0$) is interpreted as the response
to a virtual source at one of the receiver positions (${\bf x}_A$), observed by the  receiver at ${\bf x}$ (Figure \ref{Figure2}b).
\end{enumerate}
In summary, whereas in time-reversed acoustics, ${\cal G}_x({\bf x}',{\bf x}_A,-t,0)$ in equation (\ref{eqconprod1}) 
stands for the time-reversed field that is fed to sources at ${\bf x}'$ 
on ${\cal S}$ and propagated by ${\bf \nabla}'{\cal G}_x({\bf x},{\bf x}',t,0)\cdot{\bf n}$ to any ${\bf x}$ inside ${\cal S}$
(Figure \ref{Figure1}), in Green's function retrieval its reciprocal version ${\cal G}_x({\bf x}_A,{\bf x}',t,0)$ is
 temporally cross-correlated with ${\bf \nabla}'{\cal G}_x({\bf x},{\bf x}',t,0)\cdot{\bf n}$ and integrated over all ${\bf x}'$ on ${\cal S}$
(Figure \ref{Figure2}). Both approaches lead to the same homogeneous Green's function ${\cal G}_x^h({\bf x},{\bf x}_A,t,0)$ between ${\bf x}_A$ and ${\bf x}$.
\item For a homogeneous, time-variant material, the principles of time-reversed acoustics and Green's function retrieval are variations
of the homogeneous Green's function representation of equation (\ref{eq58kk}) 
for this class of material. In particular, consider the spatial convolutional product 
\begin{eqnarray}
\{\partial_{t'}  {\cal G}_t({\bf x},{\bf 0},t,t')\}*_x{\cal G}_t^a({\bf x},{\bf 0},t',t_A)\label{eqconprod2}
\end{eqnarray}
in equation (\ref{eq58kk})
 (the analysis of the other convolutional product follows the same arguments).
\begin{enumerate}
\item In time-reversed acoustics, the second Green's function (after applying the reciprocity relation 
 ${\cal G}_t^a({\bf x},{\bf 0},t',t_A)={\cal G}_{t,{\rm TR}}({\bf x},{\bf 0},-t',-t_A)$ and  a spatial convolution with a source
function) stands for the field $U_{\rm TR}({\bf x},-t')$ in a time-reversed version of the actual material. This field
(together with its sign-reversed time-derivative) is fed to sources at time instance $t'$
and propagated by the first Green's function to any time $t$ beyond $t'$, as illustrated in Figure \ref{Figure6}.
The result is the homogeneous Green's function ${\cal G}_t^h({\bf x},{\bf 0},t,t_A)$
(spatially convolved with the source function). This is interpreted as a field focusing at $t_A$ and, after having focussed,
propagating away from $t_A$ (Figure \ref{Figure6}).
\item In Green's function retrieval, the second Green's function in equation (\ref{eqconprod2}) is, 
on basis of reciprocity relation (\ref{eq1022sqshbl}) and its symmetry in ${\bf x}$,
replaced by ${\cal G}_t(-{\bf x},{\bf 0},t_A,t')$. Now the spatial convolutional product
can be interpreted as  a spatial crosscorrelation (due to the space-reversal of the second Green's function) of observations by receivers at $t$ and $t_A$,
in response to a source at $t'$, as illustrated in the left frame of Figure \ref{Figure8}. The result, after adding the other convolutional product
of equation (\ref{eq58kk}), 
is the homogeneous Green's function ${\cal G}_t^h({\bf x},{\bf 0},t,t_A)$, which (for $t>t_A$) is interpreted as the response to a virtual
source at one of the receiver times ($t_A$), observed by receivers at $t$ (the right frame of Figure \ref{Figure8}).
\end{enumerate}
In summary, whereas in time-reversed acoustics, ${\cal G}_t^a({\bf x},{\bf 0},t',t_A)={\cal G}_{t,{\rm TR}}({\bf x},{\bf 0},-t',-t_A)$ 
in equation (\ref{eqconprod2}) stands for the field in the time-reversed material 
that is fed to sources at $t'$ and 
propagated by $\partial_{t'}  {\cal G}_t({\bf x},{\bf 0},t,t')$ to any $t$ beyond $t'$ (Figure \ref{Figure6}), 
in Green's function retrieval its reciprocal version ${\cal G}_t({\bf x},{\bf 0},t_A,t')$ is
spatially cross-correlated with $\partial_{t'}  {\cal G}_t({\bf x},{\bf 0},t,t')$ (Figure \ref{Figure8}). 
Both approaches lead to the same homogeneous Green's function ${\cal G}_t^h({\bf x},{\bf 0},t,t_A)$ between $t_A$ and $t$.
\end{itemize}

\section*{Acknowledgements}
Johannes Aichele and Dirk-Jan van Manen acknowledge funding from the Swiss National Science Foundation (SNF, grant 197182). 

\appendix
\section{Analytical expressions for a homogeneous, time-invariant material}\label{AppA}

\subsection{Green's function of a homogeneous, time-invariant material}\label{AppA1}
For the special case of a  homogeneous, time-invariant material, the analytical solution of equation (\ref{eq31agbl}), with causality condition (\ref{eq31con}), reads
\begin{eqnarray}
{\cal G}_t({\bf x},{\bf x}_A,t,t_A)=
\begin{cases}
\frac{1}{4\pi\beta c^2}\frac{\delta(t-t_A- |{\bf x}-{\bf x}_A|/c)}{|{\bf x}-{\bf x}_A|},\quad&\mbox{3D},\\
&\\
\frac{1}{2\pi\beta c^2}\frac{H(t-t_A- |{\bf x}-{\bf x}_A|/c)}{\sqrt{(t-t_A)^2- |{\bf x}-{\bf x}_A|^2/c^2}},\quad&\mbox{2D},\label{eq71bl}
\end{cases}
\end{eqnarray}
where $H(t)$ is the Heaviside step function. We define the spatial Fourier transformation as
\begin{eqnarray}
\check U({\bf k}) =\int_{{\mathbb{R}}^m} U({\bf x})\exp (-i{\bf k}\cdot{\bf x}){\rm d}{\bf x},\label{eq51}
\end{eqnarray}
with $m=3$, ${\bf x}=(x,y,z)$, ${\bf k}=(k_x,k_y,k_z)$ for the 3D situation, and $m=2$, ${\bf x}=(x,z)$, ${\bf k}=(k_x,k_z)$ for the 2D situation;
$\mathbb{R}$ denotes the set of real numbers.
The inverse spatial Fourier transformation is defined as
\begin{eqnarray}
U({\bf x})=\frac{1}{(2\pi)^m} \int_{{\mathbb{R}}^m} \check U({\bf k})  \exp(i{\bf k}\cdot{\bf x}){\rm d}{\bf k}. \label{eq52}
\end{eqnarray}

In the spatial Fourier domain, the Green's function reads
\begin{eqnarray}
\check {\cal G}_t({\bf k},{\bf x}_A,t,t_A)=\exp(-i{\bf k}\cdot{\bf x}_A)H(t-t_A)\frac{\sin(|{\bf k}|c(t-t_A))}{\beta c|{\bf k}|}.\label{eq93bebl}
\end{eqnarray}
This expression holds for the 3D as well as for the 2D situation.
Analogous to equation (\ref{eq1022sqshbl}) we have for the acausal Green's function in  the spatial Fourier domain
\begin{eqnarray}
\check {\cal G}_t^a({\bf k},{\bf x}_A,t,t_A)=\check {\cal G}_t({\bf k},{\bf x}_A,t_A,t).\label{eqA9}
\end{eqnarray}

\subsection{Propagator matrix for a homogeneous, time-invariant slab}\label{AppA2}

We consider the propagator matrix ${\bf W}({\bf x},t_n,t_{n-1})$ for a homogeneous, time-invariant slab $t_{n-1}<t<t_n$, with parameters $\alpha_n$, $\beta_n$ and $c_n$.
According to equations (\ref{eq1022sqshbl}), (\ref{eq35aa}) and (\ref{eq71bl}), the upper-right element of this matrix is given by
\begin{eqnarray}\label{eqA5}
W^{U,V}({\bf x},t_n,t_{n-1})&=&{\cal G}_t^a({\bf x},{\bf 0},t_{n},t_{n-1})-{\cal G}_t({\bf x},{\bf 0},t_n,t_{n-1})\nonumber\\
&=&\begin{cases}
\frac{1}{4\pi\beta_n c_n^2}\frac{\delta(\Delta t_n+ |{\bf x}|/c)-\delta(\Delta t_n- |{\bf x}|/c)}{|{\bf x}|},\quad&\mbox{3D},\\
&\\
\frac{1}{2\pi\beta_n c_n^2}\frac{H(-\Delta t_n- |{\bf x}|/c_n)-H(\Delta t_n- |{\bf x}|/c)}{\sqrt{\Delta t_n^2- |{\bf x}|^2/c_n^2}},\quad&\mbox{2D},
\end{cases}
\end{eqnarray}
with $\Delta t_n=t_n-t_{n-1}$. The other elements of ${\bf W}({\bf x},t_n,t_{n-1})$ follow from equations (\ref{eq39})--(\ref{eq41}) and  (\ref{eqA5}).

Using equations (\ref{eq93bebl}) and (\ref{eqA9}), we obtain for the upper-right element of the propagator matrix $\check {\bf W}({\bf k},t_n,t_{n-1})$
in the spatial Fourier domain
\begin{eqnarray}\label{eqA11}
\check W^{U,V}({\bf k},t_n,t_{n-1})&=&\check {\cal G}_t^a({\bf k},{\bf 0},t_n,t_{n-1})-\check {\cal G}_t({\bf k},{\bf 0},t_n,t_{n-1}),\nonumber\\
&=&-\frac{1}{\beta_nc_n|{\bf k}|}\sin(|{\bf k}|c_n \Delta t_n).
\end{eqnarray}
This expression holds for the 3D as well as for the 2D situation.
The other elements of $\check{\bf W}({\bf k},t_n,t_{n-1})$ follow from the Fourier transforms of equations (\ref{eq39})--(\ref{eq41}) and equation (\ref{eqA11}).
Hence,
\begin{eqnarray}
\check {\bf W}({\bf k},t_n,t_{n-1})=
\begin{pmatrix}
\cos(|{\bf k}|c_n \Delta t_n) & -\frac{1}{\beta_nc_n|{\bf k}|}\sin(|{\bf k}|c_n \Delta t_n)\\
\beta_nc_n|{\bf k}|\sin(|{\bf k}|c_n \Delta t_n) & \cos(|{\bf k}|c_n \Delta t_n)
\end{pmatrix}.
\end{eqnarray}

\mbox{}\\
\centerline{{\Large References}}



\end{spacing}
\end{document}